\documentclass[11pt]{article}

\usepackage{amsmath}
\usepackage{epsfig}
\usepackage{graphics}

\newcommand{\N}{N\raise.7ex\hbox{\underline{$\circ $}}$\;$}

\begin{document}

\title{
 V.M. Red'kov\footnote{redkov@dragon.bas-net.by}\\
 Geometry of 3-Spaces with Spinor Structure
 }

\maketitle

\begin{quotation}

A special  approach to examine  spinor structure of 3-space is
proposed. It is based on the  use of the concept of a spatial
spinor defined  through taking the square root  of a real-valued
3-vector.  Two sorts of  spatial spinor according to
$P$-orientation of an initial  3-space are introduced: properly
vector or  pseudo vector one. These spinors, $\eta$ and $\xi$,
turned out to be  different functions of Cartesian coordinates. To
have a spinor space  model, you ought to use  a doubling vector
space $ \{ \; (x_{1},x_{2},x_{3} )\; \otimes \;\;
(x_{1},x_{2},x_{3})' \;\; \} $. The main idea is to develop some
mathematical technique to work with such extended models. Spinor
fields $\eta$ and $\xi$, given as functions of Cartesian
coordinates $x_{i}\oplus x_{i}'$, do not obey  Cauchy-Riemann
analyticity condition with respect to complex variable  $(x_{1} +i
x_{2}) \oplus  (x_{1} +i x_{2})'$. Spinor functions are  in
one-to-one correspondence with coordinates $x_{i}\oplus x_{i}'$
everywhere excluding the whole axis $(0,0,x_{3}) \oplus
(0,0,x_{3})'$ where they have an exponential  discontinuity. It is
proposed to consider properties  of spinor fields $\xi
(x_{i}\oplus x_{i}') $ and $\eta(x_{i}\oplus x_{i}')$ in terms of
continuity with respect to  geometrical directions in the vicinity
of   every point. The mapping of spinor field $\eta$ into $\xi$
and  inverse
 have been constructed. Two sorts of spatial spinors
are  examined with the use of curvilinear coordinates $(y_{1},y_{2},y_{3})$:
cylindrical parabolic, spherical and
parabolic ones.  Transition from vector to spinor models is achieved by doubling
initial parameterizing domain $G(y_{1},y_{2},y_{3})  \; \Longrightarrow
\tilde{G}(y_{1},y_{2},y_{3})$ with new identification rules  on the  boundaries.
Different spinor space  models are built on explicitly different spinor
fields $\xi(y)$ and $\eta(y)$. Explicit form of the mapping  spinor field $\eta(y)$
of pseudo vector  model into spinor $\xi(y)$ of properly vector one is given,
it contains explicitly complex conjugation.

\end{quotation}

\section{Introduction}

\hspace{7mm}
In the  literature, the  problem of the  so-called spinor
structure of physical space-time was
extensively  discussed [1-40].  There were  considered
 both  possible  experimental tests and
mathematical methods  to describe  such  a structure (see also [41-44]).

The main idea of the  present treatment  is
to elaborate  certain  approach to this  problem
in the frame of  mathematical technique, simple  and natural as much as  possible,
for physicists without  refined knowledge in  contemporary  topology and  geometry.
In other words, the  idea is to show that
 old and  naive mathematical tools based on the
use of explicit  coordinate language, which  is yet of the  most significance in any
experiments-oriented  physics,
might be  quite  sufficient  to describe
adequately subtleties and peculiarities associated
with  possible spinor structure of  physical space-time.

For simplicity,  this work is  restricted to  a  "non-relativistic"\hspace{2mm}
spinor  model only when
2-component spinors  of the  unitary group $SU(2)$  are taken into consideration.
Brief preliminary  remarks should be  given
of the  concept of spatial spinor  -- primary mathematical
object associated with a "point of a spinor space". We will start with the well-known
  Cartan's classification
of 2-spinors with respect to spinor $P$-reflection: namely,  the simplest irreducible
representations of the  unitary extended group

\begin{eqnarray}
\tilde{SU}(2) =  \Big\{\;\; g \in SU(2) \oplus J = \left (
\begin{array}{cc} i & 0 \\ 0  & i \end{array} \right ) , \;
   \det
\; g = +1, \; \det \; J = -1 \;\;  \Big\} \qquad \label{1.1}
\end{eqnarray}

\noindent are 2-component spinors of two types $T_{A}$
\begin{eqnarray}
T_{1}: \qquad  T_{1}(g) = g, \; T_{1}(J) = + J \; ,
\nonumber
\\
T_{2}: \qquad  T_{2}(g) = g, \; T_{2}(J) = - J \; .
\label{1.2}
\end{eqnarray}

 With this in mind, there are two ways to
 construct 3-vector (complex-valued in general)
in terms of 2-spinors
\begin{eqnarray}
1. \qquad (\xi \otimes \xi ^{*}) =
 a \;+\; a_{j} \; \sigma ^{j}\; \; , \;\;
a  = \sqrt{ a_{j} \; a_{j}} \; ,
\nonumber
\\
2.  \qquad \qquad \qquad \;\;\;
(\eta \otimes \eta ) =  \;( c_{j} \;+\; i \; b_{j}) \;\sigma ^{j}\;
\; .
\label{1.3}
\end{eqnarray}

From (\ref{1.3}) it follows that when  spinor $\eta$ is
either of the type 1 or of the  type 2, real-valued 3-vector
$a_{j}$  is  a  pseudo  vector. In turn, when
 $\xi$ is either of the type 1 or of the  type 2, real-valued 3-vectors
$c_{j}$ and $b_{j}$
are both   proper   vectors.
Evidently, variant 1 provides us  with possibility
to build a spinor model for  pseudo vector
3-space $\tilde{\Pi}_{3}$, whereas variant 2 leads to a spinor model of  properly
vector 3-space  $\tilde{E}_{3}$.
In other words, according to which way of taking
the square root of three real numbers --
components  of a 3-vector $x_{i}$, one will arrive at two different spatial spinors
\begin{eqnarray}
\xi \;  \Longleftrightarrow  \; a_{j} \; , \qquad
\eta\;  \Longleftrightarrow  \; c_{j} \;\; \mbox{or} \;\; (b_{j}) \; .
\label{1.4}
\end{eqnarray}

These spinors, $\eta$ and $\xi$ respectively,
turned out to be  different functions of
Cartesian coordinates  $(x_{1},x_{2},x_{3})$.
  Evidently, to have in hand a spinor space  model,
you are  to use in a sense  a doubling vector space
\begin{eqnarray}
\{ \; (x_{1},x_{2},x_{3} )\; \otimes \;\; (x_{1},x_{2},x_{3})' \;\; \}.
\label{1.5}
\end{eqnarray}

The main idea is to develop some  mathematical technique
to work with such extended models.
Spinor fields $\eta$ and $\xi$, constructed as functions of Cartesian
coordinates $x_{i}\oplus x_{i}'$, do not obey  Cauchy-Riemann analyticity condition
with respect to complex variable  $(x_{1} + ix_{2})$.
Spinor functions are in one-to-one correspon\-dence with coordinates
$x_{i}\oplus x_{i}'$ everywhere excluding
 the  whole axis $(0,0,x_{3}) \oplus (0,0,x_{3})'$
they have an exponential and discrete
 $\pm$-sign discontinuities. After extending models to
spinor ones only exponential discontinuity remains. It was  proposed to consider
properties  of spinor fields in terms
of continuity with respect to  geometrical directions
in the vicinity of the  every point.

In addition,  two sorts of spatial spinors depending on  $P$-orientation
are examined with the use of curvilinear coordinates $(y_{1},y_{2},y_{3})$.
Transition from vector to spinor models is achieved by doubling
initial parameterizing domain: $G(y_{1},y_{2},y_{3})  \; \Longrightarrow
\tilde{G}(y_{1},y_{2},y_{3})$ with new identification rules  on the  boundary.
Different spinor space  models are built on explicitly different spinor
fields $\xi$ and $\eta$. Explicit form of the mapping  spinor field $\eta(y)$
of pseudo vector  model into spinor $\xi(y)$ of properly vector one is given,
it contain explicitly complex conjugation.
Three most commonly used coordinate
systems --   spherical, parabolic ones,
and cylindrical parabolic --  have been considered
in detail.

\section{Pseudo vector space $\Pi_{3}$  and  its spatial spinor  $\xi $}

\hspace{7mm}
Let  $\xi $ be either a spinor of the first or second type,
then a conjugate spinor
$\xi^{*}$  will be of the second or first type respectively.
Combining them into a 2-rank spinor,
we get a pseudo scalar $a$ and pseudo vector $a_{j}$
\begin{eqnarray}
(\xi  \otimes \xi ^{*}) =
 a + a_{j}  \sigma ^{j} \; ,\;
a^{(J)} = + a\; , \;
a^{(J)}_{j} = + a_{j} \; .
\label{2.1a}
\end{eqnarray}

\noindent Involved quantities transform under $SU(2)$ according to
(the notation is  used $(\vec{n}^{\;\times} ) _{ij} = \epsilon_{ijk} n_{k}$)
\begin{eqnarray}
\xi' = B(n)  \xi  , \qquad
B(n) = I n_{0} - i \sigma^{j}n_{j} \;  , \qquad
\label{2.1b}
\\
a'_{j} = 0_{jl}(n)  a_{l} \; , \; 0(n) =   I +  2\; [n_{0} \vec{n}^{\times } +
(\vec{n}^{\;\times })^{2}]  \;\;  .
\nonumber
\end{eqnarray}

\noindent The task is to find an explicit form of $a$  and $a_{j}$ in terms of
spinor components. With the  notation
\begin{eqnarray}
\xi  = \left ( \begin{array}{c}
x \\ y \end{array} \right )  \; ,   \;\;\;
(\xi  \otimes  \xi^{*} ) = \left ( \begin{array}{cc}
x x^{*}  &  x y^{*}  \\  y x^{*}   &  y y^{*}
\end{array} \right ) \; ,
\label{2.2a}
\end{eqnarray}

\noindent we have
\begin{eqnarray}
a_{1} = {1 \over 2}\; (y x^{*} + x y^{*})\; , \;\;
a_{2} = {i \over 2}\; (x y^{*} - x^{*} y)\; ,
\nonumber
\\
a_{3} = {1 \over 2} \;(x x^{*} - y y^{*})\; ,\;\;
a = {1 \over 2} \; (x x^{*} + y y^{*})\; .
\label{2.2b}
\end{eqnarray}

\noindent Observing identity
$\vec{a}^{\;2} = {1 \over 4} (x x^{*} + y y^{*} )^{2}$, one  concludes that
the scalar  $a$ is a positive square root of $\vec{a}^{\;2}$:
$
\;\; a = + \sqrt{\vec{a}^{\;2}}   \;.
$

Needless to say that multiplying an initial spinor
 $\xi $  by a phase factor  $e^{i\alpha }$
does not affect both  $a$ and $a_{j}$;  this peculiarity will find its corollary in
finding a spinor $\xi$ from a  given vector $a_{j}$.
Now  we are ready to invert relations
(\ref{2.2b}). To this end one should take $\eta$
in a special form ($N , M \in  [ 0 , \infty  )$)
\begin{eqnarray}
\xi  = \left ( \begin{array}{c}
N  e^{i n} \\ M  e^{ i m}   \end{array} \right )  , \qquad
n , m \in  [-\pi,+\pi] \; .
\label{2.3a}
\end{eqnarray}

\noindent  Substituting (\ref{2.3a})  into  (\ref{2.2b}), one  gets
\begin{eqnarray}
a_{1} \pm i a_{2} = N M \;  e^{\pm i (m - n)} \; ,
\nonumber
\\
a_{3} = {1 \over 2} ( N^{2} - M^{2})  ,\;\;
a = {1 \over 2} ( N^{2} + M^{2}) \;.
\label{2.3b}
\end{eqnarray}

\noindent From (\ref{2.3b}) one can see that components $a_{1}$ and $a_{2}$
determine only the difference $(m - n)$. In turn,  $a_{3}$  and   $a$
will fix  $M$ and $N$:

\vspace{10mm}

\hspace{-3mm}
\unitlength=0.5mm
\begin{picture}(40,50)(-100,-10)
\special{em:linewidth 0.4pt}
\linethickness{0.4pt}
\put(0,0){\vector(1,0){40}}
\put(0,0){\vector(0,1){40}}
\put(43,-5){$M$}
\put(-10,40){$N$}
\put(0,0){\line(1,1){35}}
\end{picture}
\begin{center}
FIG.1. $(M,N)$-diagram
\end{center}

\noindent Here the line $N = M$  corresponds to the  plane $a_{3 } = 0$;
sub-set $ N > M$  refers to upper half-space
$\Pi ^{+}_{3} ( a_{3} > 0 )$; and sub-set $N < M  $ refers to lower half-space
$\Pi ^{-}_{3} ( a_{3} < 0 )$; $M = 0  $  refers to half-axis  $a_{3} > 0$;
$N = 0  $  refers  to half-axis  $a_{3} < 0$; initial point  $(0,0,0)$ is  given by
 $\xi = 0$.
 For different regions of the $\Pi$-space the following designation
 will be used (see Fig. 2).

\vspace{+3mm}

\unitlength=0.5 mm
\begin{picture}(100,95)(-100,-50)
\special{em:linewidth 0.4pt}
\linethickness{0.4pt}

\put(-40,0){\vector(1,0){80}}
\put(0,-40){\vector(0,1){80}}
\put(+30,+30){\vector(-1,-1){60}}
\put(-37,-30){$a_{1}$}
\put(+40,-5){$a_{2}$}
\put(-9,+40){$a_{3}$}
\put(-30,+20){\vector(+1,0){28}}
\put(-35,+13){$\Pi^{+}_{0}$}
\put(+30,-20){\vector(-1,0){28}}
\put(+35,-20){$\Pi^{-}_{0}$}
\put(+10,+30){$\Pi^{+}$}   \put(-12,-30){$\Pi^{-}$}
\end{picture}

\begin{center}
FIG. 2. $\Pi$-space regions
\end{center}


Instead of the  variables  $n$ and $m$ it is useful to take  two new ones
$\gamma $ and  $\kappa$:
\begin{eqnarray}
\kappa  = ( m + n ) \; , \; \gamma = ( m - n ) \; ,
\nonumber
\\
n = {1 \over 2} ( \kappa  - \gamma ) \; , \;\;
m = {1 \over 2} ( \kappa  + \gamma )  \;  .
\label{2.4}
\end{eqnarray}

\noindent Correspondingly, the  domain  $G(n,m)$,
a square centered in $(0,0)$ and with area
$(4\pi ^{2})$, will change into a rhombus with area $(8\pi ^{2})$:

\unitlength=0.5 mm
\begin{picture}(160,100)(-60,0)
\special{em:linewidth 0.4pt}
\linethickness{0.4pt}
\put(+5,+50){\vector(+1,0){50}}    \put(+52,+43){$n$}
\put(+30,+25){\vector(0,+1){50}}   \put(+22,+75){$m$}
\put(+15,+35){\line(+1,0){30}}
\put(+15,+35){\line(0,+1){30}}
\put(+45,+65){\line(-1,0){30}}
\put(+45,+65){\line(0,-1){30}}

\put(+3,+45){$-\pi$}    \put(+48,+53){$+\pi$}
\put(+20,+28){$-\pi$}    \put(+33,+67){$+\pi$}
\put(+70,+50){\vector(+1,0){80}}  \put(+150,+45){$\gamma$}
\put(+110,+10){\vector(0,+1){80}}  \put(+105,+90){$\kappa$}
\put(+80,+50){\line(+1,+1){30}}
\put(+80,+50){\line(+1,-1){30}}
\put(+140,+50){\line(-1,+1){30}}
\put(+140,+50){\line(-1,-1){30}}
\put(+140,52){$+2\pi$}  \put(68,+44){$-2\pi$}
\put(+93,+17){$-2\pi$}  \put(+111,+80){$+2\pi$}
\end{picture}

\vspace{-5mm}

\begin{center}
FIG. 3. $(\gamma, k)$-diagram
\end{center}

\noindent
In the  variables $(\kappa ; N, M, \gamma )$,
spinor $\xi$ looks as (take  note on a
phase factor $e^{ik/2}$)
\begin{eqnarray}
\xi = e^{i\kappa /2} \left ( \begin{array}{c}
N\; e^{-i\gamma /2}  \\ M \; e^{+i\gamma /2} \end{array} \right )\; ,
\label{2.5a}
\end{eqnarray}

\noindent and eqs. (\ref{2.3b}) will read
\begin{eqnarray}
a_{1} \pm i a_{2}  = N M \; e^{\pm i\gamma } \;, \;\; \qquad
\nonumber
\\
a_{3} = {1 \over 2} ( N^{2} - M^{2}) \; ,\;\;
a = {1 \over 2} ( N^{2} + M^{2}) \;  .
\label{2.5b}
\end{eqnarray}

\noindent
One should note that the variable  $\kappa $ does not enter (\ref{2.5b}).
Besides, ranging the variable  $\gamma $
in the  interval  $[- 2\pi ,\; + 2\pi  ]$
(see Fig. 3) ensures required double covering
of the ordinary plane $(a_{1}, a_{2})$.
In other words, three parameters  $(M, N, \gamma )$  are sufficient to parameterize
spinor model $\tilde{\Pi}_{3}$  built upon a pseudo vector space $\Pi_{3}$.
To this model $\tilde{\Pi}_{3}$ there is a corresponded
spinor field $\xi (\vec{a})$ (in the
following,  the  factor $e^{ik/2}$ will be  omitted)
\begin{eqnarray}
\xi  =  \left ( \begin{array}{c}
\sqrt{a + a_{3} }  e^{-i\gamma/2}  \\
\sqrt{a - a_{3}}  e^{+i\gamma /2}  \end{array} \right )
 , \;\;
e^{i\gamma } = { a_{1} + i a_{2} \over
\sqrt{ a^{2}_{1} + a^{2}_{2} } } \; .
\label{2.6}
\end{eqnarray}

It should be noted that in describing  $\Pi^{+}_{0}$ and $\Pi^{-}_{0}$ there arise
peculiarities:  at the  whole axis  $a_{3}$ eqs. (\ref{2.6}) contain ambiguity
$(0+i0)/0$ (and expressions for $\xi$ will contain a mute  angle variable $\Gamma: \;
 \gamma  \rightarrow \Gamma$ )
\begin{eqnarray}
\Pi^{+}_{0}\;\; : \qquad \xi ^{+}_{0} =
\left ( \begin{array}{c}
\sqrt{+2a_{3}} \; e^{-i\Gamma /2} \\ 0 \end{array} \right ) \; ,
\label{2.7a}
\\
\Pi^{-}_{0}\;\; :  \qquad
 \xi ^{-}_{0}  = \left ( \begin{array}{c}
 0  \\ \sqrt{-2a_{3} }\; e^{+i\Gamma /2} \end{array} \right ) \;,
\label{2.7b}
\\
e^{i\Gamma }  =  \lim_{a_{1},a_{2} \rightarrow 0}
{ a_{1} + i a_{2} \over \sqrt{ a^{2}_{1} + a^{2}_{2} }}\; .
\label{2.7c}  \qquad \qquad
\end{eqnarray}

\noindent
At the  plane  $a_{3}= 0$,  spinor $\xi$ reads as
\begin{eqnarray}
\xi  =  \left ( \begin{array}{c}
\sqrt{ a^{2}_{1} + a^{2}_{2} }\;  e^{-i\gamma /2} \\[2mm]
\sqrt{a^{2}_{1} + a^{2}_{2} } \;  e^{+i\gamma /2}
\end{array}  \right ) \; .
\label{2.8}
\end{eqnarray}

\section{Proper vector space $E_{3}$ and its
spinor model $\tilde{E}_{3}$}

\hspace{8mm}
In this Section we are going in the same line
to define a spatial spinor associated with a
proper vector space $E_{3}$.
Let  $\eta $ be a spinor  either  of the first or second type.
It leads  to a 2-rank spinor $(\eta \otimes  \eta )$,
equivalent to  a couple of real-valued proper vectors
 $c_{j}$ Ё $b_{j}$:
\begin{eqnarray}
(\eta \otimes \eta ) = \;( c_{j} + i\; b_{j} ) \sigma^{j}\;\; .
\label{3.1a}
\end{eqnarray}

\noindent With respect to  $J$-reflection,  involved quantities
$\eta $ and $(c_{j},\; b_{j})$  are transformed according to
\begin{eqnarray}
\eta ^{(J)} = +i \eta  \;\; (or \; -i\eta )\;, \qquad
c^{(J)}_{j} = - c_{j}  \;,  \;\; b^{(J)}_{j} = - b_{j}\; ,
\label{3.1b}
\end{eqnarray}

\noindent and under continuous group $SU(2)$
\begin{eqnarray}
\eta' = B(n)  \eta \; , \; c '_{i} = 0_{ij}(n) c_{j} \;  ,\;
b'_{i} = 0_{ij}(n) b_{j} \; .
\label{3.1c}
\end{eqnarray}

The task is to  find  vectors $\vec{c}$ and  $\vec{b}$
in terms  of spinor $\eta$ components.
With  the  same  notation
\begin{eqnarray}
\eta  = \left ( \begin{array}{c}
N  e^{i n}  \\  M  e^{ i m}
\end{array} \right )  \;
\label{3.2a}
\end{eqnarray}

\noindent
after simple calculating we get
\begin{eqnarray}
\left. \begin{array}{l}
c_{1} = {1 \over 2} (- M^{2} \sin  2m + N^{2} \sin 2n ) \; , \\[1mm]
c_{2} = {1 \over 2} (+ M^{2} \cos 2m +  N^{2} \cos 2n ) \;, \\[1mm]
c_{3} = - M N \sin (m + n) \; , \\[1mm]
b_{1} = {1 \over 2} ( M^{2} \cos 2m - N^{2} \cos 2n ) \; , \\[1mm]
b_{2} = {1 \over 2} ( M^{2} \sin  2m + N^{2} \sin  2n )\; , \\[1mm]
b_{3} = + M N \cos  (m+n) \;.
\end{array} \right.
\label{3.2b}
\end{eqnarray}

\noindent
Again, instead of   $(m,n)$  we will use  $(\kappa ,\gamma )$
(see  (\ref{2.4})), then eqs. (\ref{3.2b}) read as
\begin{eqnarray}
\vec{c} = \vec{e}_{f} \cos \kappa - \vec{f} \sin \kappa \; ,\qquad
 \vec{b} =  \vec{e}_{f} \sin \kappa + \vec{f} \cos \kappa \; ,
\label{3.3a}
\end{eqnarray}

\noindent where $\vec{f}$ and  $\vec{e}_{f}$  are given  by
\begin{eqnarray}
\vec{f} = \left ( \begin{array}{c}
{1 \over 2} (M^{2} - N^{2})  \cos  \gamma  \\[1mm]
{1 \over 2} (M^{2} - N^{2})  \sin  \gamma  \\ [1mm]
+M N \end{array} \right )  ,
\vec{e}_{f} =  \left (  \begin{array}{c}
- f \sin \gamma   \\[1mm]  + f \cos \gamma  \\ [1mm] 0
\end{array} \right )  .
\nonumber
\end{eqnarray}

\noindent All four vectors  $(\vec{f}, \; \vec{e}_{f}, \;
\vec{c}, \; \vec{b})$ have the same length
\begin{eqnarray}
\mid \vec{f} \mid  = \mid  \vec{e}_{f} \mid  =
\mid  \vec{c} \mid  = \mid \vec{b}\mid    =
(M^{2} + N^{2}) / 2 \; .
\nonumber
\end{eqnarray}

\noindent Besides, two orthogonality conditions
$\;
\vec{f} \; \vec{e}_{f}  = 0 $ and $
\vec{b} \; \vec{c}  = 0 \;$ hold.

Now, we are at the  point to determine
certain sub-set of spinors  in  $\eta(\kappa ;\; N, M, \gamma )$,
which could  be  suitable to parameterize  correctly spinor space
$\tilde{E}_{3}$, covering twice an initial vector space $E_{3}$.

Starting from  the set $\; (\kappa  = 0;\; N ,M , \gamma  ) \;$
and respective  sets of
vectors $\vec{c}$ and $\vec{b}$ :
\begin{eqnarray}
\vec{b} = + \vec{f}  , \;\;\;
\vec{c} = + \vec{e}_{f} \;  , \qquad
\vec{f} = \left ( \begin{array}{c}
{1 \over 2} (M^{2} - N^{2})  \cos  \gamma  \\[1mm]
{1 \over 2} (M^{2} - N^{2})  \sin  \gamma  \\ +N N
\end{array}  \right )       ,
\label{3.5a}
\end{eqnarray}

\noindent and demanding   parameters $M,N,\gamma $
to be  ranged as follows
\begin{eqnarray}
M > N > 0 \; , \; \gamma  \in  [- 2\pi , \; + 2\pi ]\; .
\label{3.5b}
\end{eqnarray}

\noindent  Vector  $\vec{b}$ covers upper half-space  $E^{+}_{3}$ twice;
respective  spinor  $\eta^{+}$   looks as
\begin{eqnarray}
\eta ^{+} = \left ( \begin{array}{c}
\sqrt{b - (b^{2}_{1} + b^{2}_{2})^{1/2} } \; e^{-i\gamma /2} \\[2mm]
\sqrt{b - (b^{2}_{1} + b^{2}_{2})^{1/2} } \; e^{+i\gamma /2}
\end{array} \right )  \; ,\qquad
e^{i\gamma } = { b_{1}+ i b_{2} \over
\sqrt{ b^{2}_{1} + b^{2}_{2} }} \;. \qquad \qquad
\label{3.5c}
\end{eqnarray}

Now let us start with the sub-set
$
(\kappa  = \pi ; N , M , \gamma  )
$
 and respective vectors
$\vec{c}$ and  $\vec{b}$:
\begin{eqnarray}
\vec{b} = -  \vec{f}  ,  \;\;  \vec{c} = -  \vec{e}_{f}\; ,
\qquad
\vec{f} = \left ( \begin{array}{c}
{1 \over 2} (M^{2} - N^{2}) \; \cos  \gamma  \\[1mm]
{1 \over 2} (M^{2} - N^{2}) \; \sin  \gamma  \\  + N N
\end{array} \right ) \; .
\label{3.6a}
\end{eqnarray}

\noindent If one again expects the parameters  $M, N, \gamma $
to vary according to (\ref{3.5b}),
then the vector  $\vec{b}$ in (\ref{3.6a}) covers
a  lower half-space $E^{-}_{3}$ twice;
expression for spinor $\eta ^{-}$  looks as
\begin{eqnarray}
\eta ^{-} = i  \left ( \begin{array}{c}
\sqrt{ b - (b^{2}_{1} + b^{2}_{2})^{1/2}} \; e^{-i\gamma /2} \\[2mm]
\sqrt{ b - (b^{2}_{1} + b^{2}_{2})^{1/2}} \; e^{+i\gamma /2}
\end{array} \right ) \; ,  \;\;
e^{+i\gamma /2}  = - i \sqrt{{b_{1} + i b_{2} \over
\sqrt{b^{2}_{1} + b^{2}_{2}}} }\; , \qquad \qquad
\label{3.6b}
\end{eqnarray}

\noindent or in equivalent form
\begin{eqnarray}
\eta ^{-} =  \left ( \begin{array}{ll}
\sqrt{b - (b^{2}_{1}+b^{2}_{2})^{1/2} } &
\left [ -\sqrt{{b_{1}+ib_{2} \over (b^{2}_{1}+b^{2}_{2})^1/2}} \;\right ]^{*} \\[3mm]
\sqrt{b + (b^{2}_{1}+b^{2}_{2})^{1/2} } &
\left [ +\sqrt{ {b_{1}+ ib_{2} \over (b^{2}_{1} + b^{2}_{2})^1/2}}
\; \right ] \end{array} \right ) .
\label{3.6c}
\end{eqnarray}

It is natural to expect a spinor field $\eta$ to be continuous at the
plane  $b_{3} = 0$, for  this one
must use in  (\ref{3.6b}) and (\ref{3.5c})   the same
square root of  $(b_{1} + i b_{2})$. Thus,  spinor $\eta ^{+\cap -}$  reads
\begin{eqnarray}
\eta ^{+\cap -} = \left ( \begin{array}{c}
0 \\ \sqrt{2 (b_{1} + i \; b_{2}) }
\end{array} \right ) \; .
\label{3.7}
\end{eqnarray}

Else one  point should be clarified.
Above, two variables    $m$ and $n$ were taken  as independent, each of them
varies in the interval $[-\pi , \;+ \pi ]$. As a result,
alternative variables  $(\gamma,\kappa )$
change inside the rhombus
$G(\gamma, \kappa  )$ with area $8\pi ^{2}$ (see Fig. 3).

In accordance with this,  the variable  $\gamma \in G(\gamma,\kappa = 0)$
will  lie  automatically  inside the  interval $[-2\pi, +2\pi ] $.
 It is  just we need to parameterize spinor half-space.
In turn, $\gamma \in G(\gamma, \kappa = \pi /2)$ lies  only in the
interval $[-\pi, + \pi ]$. However,  to parameterize  spinor half-space we need that
the  variable $\gamma \in [-2\pi, + 2 \pi ]$.

There is no contradiction here because
the  domain $G(n,m)$ is  equivalent to  both the domain $G(\kappa, \gamma)$
 mentioned above  and
another domain $G'(\gamma, \kappa )$
(identification rules of the boundary points see in diagrams
below)



\begin{picture}(160,100)(-60,+10)
\special{em:linewidth 0.4pt}
\linethickness{0.4pt}
\unitlength=0.5 mm

\put(+20,+90){$G(n,m)$}
\put(+10,+50){\vector(+1,0){50}}    \put(+50,+45){$n$}
\put(+30,+25){\vector(0,+1){50}}   \put(+25,+75){$m$}
\put(+15,+35){\line(+1,0){30}}
\put(+15,+35){\line(0,+1){30}}
\put(+45,+65){\line(-1,0){30}}
\put(+45,+65){\line(0,-1){30}}
\put(+6,+35){$A''$}    \put(+47,+35){$A'''$}
\put(+8,+65){$A'$}     \put(+47,+65){$A$}

\put(+80,+90){$G(\gamma,\kappa)$}
\put(+75,+50){\vector(+1,0){75}}  \put(+145,+43){$\gamma$}
\put(+110,+10){\vector(0,+1){80}}  \put(+115,+90){$\kappa$}
\put(+80,+50){\line(+1,+1){30}}
\put(+80,+50){\line(+1,-1){30}}
\put(+140,+50){\line(-1,+1){30}}
\put(+140,+50){\line(-1,-1){30}}
\put(+140,52){$A'''$}  \put(70,+43){$A'$}
\put(+98,+16){$A''$}  \put(+112,+80){$A$}
\put(15,40){\line(1,0){30}}
\put(15,45){\line(1,0){30}}
\put(15,55){\line(1,0){30}}
\put(15,60){\line(1,0){30}}
\put(20,35){\line(0,1){30}}
\put(25,35){\line(0,1){30}}
\put(35,35){\line(0,1){30}}
\put(40,35){\line(0,1){30}}
\put(105,25){\line(+1,+1){30}}
\put(100,30){\line(+1,+1){30}}
\put(95,35){\line(+1,+1){30}}
\put(90,40){\line(+1,+1){30}}
\put(85,45){\line(+1,+1){30}}
\put(85,55){\line(+1,-1){30}}
\put(90,60){\line(+1,-1){30}}
\put(95,65){\line(+1,-1){30}}
\put(100,70){\line(+1,-1){30}}
\put(105,75){\line(+1,-1){30}}
\end{picture}


\begin{center}
FIG.4. $G(\gamma,\kappa)$-diagram
\end{center}   \vspace{-10mm}

\vspace{+5mm}
\unitlength=0.6 mm
\begin{picture}(100,70)(-100,-30)
\special{em:linewidth 0.4pt}
\linethickness{0.4pt}

\put(-40,+20){$G'(\gamma, \kappa)$}
\put(-40,0){\vector(+1,0){80}} \put(+40,-5){$\gamma$}
\put(+33,+3){$+2\pi$}    \put(-42,-5){$-2\pi$}
\put(0,-25){\vector(0,+1){50}}   \put(+3,+25){$\kappa$}
\put(-9,-20){$-\pi$}    \put(+3,+18){$+\pi$}

\put(-30,-15){\line(+1,0){60}}
\put(-30,+15){\line(+1,0){60}}
\put(-30,-15){\line(0,+1){30}}
\put(+30,-15){\line(0,+1){30}}

\put(-30,-10){\line(+1,0){60}}
\put(-30,-5){\line(+1,0){60}}
\put(-30,+5){\line(+1,0){60}}
\put(-30,+10){\line(+1,0){60}}

\put(-30, +15){\line(+1,-1){30}}
\put(-20, +15){\line(+1,-1){30}}
\put(-10, +15){\line(+1,-1){30}}
\put(0, +15){\line(+1,-1){30}}

\put(0, +15){\line(-1,-1){30}}
\put(+10, +15){\line(-1,-1){30}}
\put(+20, +15){\line(-1,-1){30}}
\put(+30, +15){\line(-1,-1){30}}
\end{picture}

\begin{center}
FIG. 5. $G'(\gamma,\kappa)$-diagram
\end{center}

\vspace{5mm}

\vspace{5mm}
\noindent Transition from $G(\kappa ,\gamma )$  to  $G'(\kappa ,\gamma )$
can be additionally explained by the  diagram

\vspace{5mm}

\unitlength=0.46 mm
\begin{picture}(170,90)(-50,0)
\special{em:linewidth 0.4pt}
\linethickness{0.4pt}
\put(+10,+85){$G(\gamma, \kappa)$}
\put(+5,40){\vector(+1,0){75}}    \put(+75,+35){$\gamma$}
\put(40,0){\vector(0,+1){80}}    \put(+45,+80){$\kappa$}
\put(+10,+40){\line(+1,+1){30}}
\put(+10,+40){\line(+1,-1){30}}
\put(+70,+40){\line(-1,+1){30}}
\put(+70,+40){\line(-1,-1){30}}
\put(+25,+55){\line(+1,0){30}}
\put(+25,+25){\line(+1,0){30}}
\put(+44,+58){$1$} \put(+35,58){$2$}
\put(+44,+18){$3$} \put(+35,18){$4$}

\put(+100,+85){$G'(\gamma, \kappa)$}
\put(+90,+40){\vector(+1,0){80}}  \put(+170,+35){$\gamma$}
\put(130,0){\vector(0,+1){80}}  \put(+135,+80){$\kappa$}

\put(+100,+25){\line(+1,0){60}}
\put(+100,+55){\line(+1,0){60}}
\put(+100,+25){\line(0,+1){30}}
\put(+160,+25){\line(0,+1){30}}

\put(+100,+40){\line(+1,+1){15}}  \put(+104,+48){$3$}
\put(+100,+40){\line(+1,-1){15}}  \put(+104,+28){$1$}
\put(+160,+40){\line(-1,+1){15}}  \put(+155,+48){$4$}
\put(+160,+40){\line(-1,-1){15}}  \put(+155,+28){$2$}
\end{picture}

\begin{center}
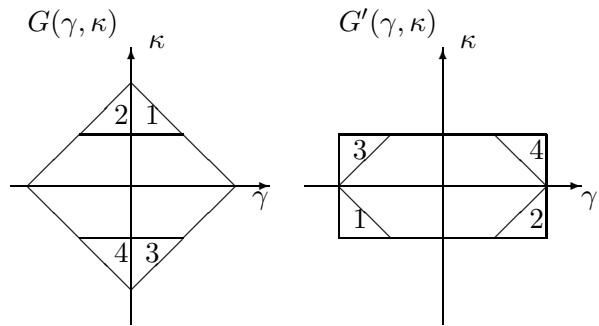
FIG. 6. Transition from  $G(\gamma, \kappa)$ to $G'(\gamma, \kappa)$
\end{center}

\vspace{5mm}
In closing, let us dwell on peculiarities in parameterizing  subsets
 $\tilde{E}^{+}_{0}$ and  $\tilde{E}^{-}_{0}$ by spinor field $\eta$.
As for  $\tilde{E}^{+}_{0}$ we  have
\begin{eqnarray}
\vec{b} = +\vec{f}\;,\;\;  \vec{c} = +\vec{e}_{f} \; , \;\;
\vec{f} = \left ( \begin{array}{c}
0 \\ 0 \\ + M N        \end{array} \right )    \; .
\label{3.8a}
\end{eqnarray}

\noindent That is  $M = N$ and the  variable  $\gamma $  is  mute, therefore
\begin{eqnarray}
\eta ^{(+)}_{0} = \sqrt{+b_{3} }
\left ( \begin{array}{c}
e^{-i\Gamma /2} \\ e^{+i\Gamma /2} \end{array} \right ) \; , \qquad
e^{i\Gamma } =
\lim_{b_{1} \rightarrow 0 , b_{2} \rightarrow 0}
{ b_{1} + i b_{2} \over \sqrt{b^{2}_{1} + b^{2}_{2} }}  \; .
\label{3.8b}
\end{eqnarray}

\noindent Analogously, for  $\tilde{E}^{+}_{0}$ we have
\begin{eqnarray}
\vec{b} = -\vec{f}\; , \qquad  \vec{c} = -\vec{e}_{f} \; ,\qquad
\vec{f} = \left ( \begin{array}{c}
 0 \\ 0 \\ + M N \end{array} \right ) \; ,
\label{3.9a}
\\
\eta ^{(-)}_{0} = \sqrt{-b_{3} } \;
\left ( \begin{array}{c}
-e^{-i\Gamma /2} \\ + e^{+i\Gamma /2}  \end{array} \right ) \; ,
\qquad
 e^{i\Gamma } =
\lim_{b_{1} \rightarrow 0 , b_{2} \rightarrow 0 }
{b_{1}+ i b_{2} \over \sqrt{b^{2}_{1} + b^{2}_{2} }} \; .
\label{3.9b}
\end{eqnarray}

\section{Spatial spinor $\xi _{a_{3}} (a_{1}+ia_{2})$
and Cauchy-Riemann analitycity}

It is  natural to consider two components of spinor  field
 $\xi  = \xi (a_{j})$ as
complex-valued functions of  $z = a_{1} + i a_{2}$  and
a  real-valued coordinate  $a_{3}$:
\begin{eqnarray}
\xi  = \left ( \begin{array}{c}
\xi ^{1}_{a_{3}} (a_{1}+ i a_{2}) \\[2mm]
 \xi ^{2}_{a_{3}}(a_{1}+ i a_{2})
\end{array} \right ) \; .
\label{4.1}
\end{eqnarray}

\noindent Since spinor components depend   upon $a_{1} +i a_{2}$ and  its conjugate
$a_{1} -i a_{2}$, they  do not differentiable in Cauchy-Riemann sense.
Let us  enlarge on the subject. Cauchy-Riemann (C-R) condition has the form
\begin{eqnarray}
 z = x + i y , \;\; f(z) = U  + i\; V \; \; , \;\;
{\partial U \over \partial x} - {\partial V \over \partial y} = 0 \; ,\;
{\partial U \over \partial y} + {\partial V \over \partial x} = 0\; .
\label{4.2}
\end{eqnarray}

\noindent For  spinor components  it will be convenient to use the notation
\begin{eqnarray}
\xi ^{1} = f^{+} ( \cos {\gamma\over 2}   - i \sin {\gamma\over 2}   ) =  U^{1} + i V^{1} \; , \;\;
\nonumber
\\
\xi ^{2} = f^{-} ( \cos {\gamma \over 2}   + i \sin {\gamma \over 2}   ) =  U^{2} + i V^{2} \; , \;\;\;
\label{4.3a}
\end{eqnarray}

\noindent where
\begin{eqnarray}
f^{\pm} = \sqrt{a \pm a_{3} } \; , \;
e^{+i\gamma /2} = \cos { \gamma \over 2}  +  i   \sin { \gamma \over 2}  \; .
\label{4.3b}
\end{eqnarray}

\noindent The formulas will be  needed:
$$
{\partial f^{+} \over \partial a_{1}} = {a_{1} \over 2a \sqrt{a + a_{3}}} \; ,  \;
{\partial f^{+} \over \partial a_{2}} = {a_{2} \over 2a \sqrt{a + a_{3}}} \; ,
$$
$$
{\partial f^{-} \over \partial a_{1}}  = {a_{1} \over 2a \sqrt{a-a_{3}}} \; ,\;
{\partial f^{-} \over \partial a_{2}}  = {a_{2} \over 2a \sqrt{a - a_{3}}} \; ,
$$

\noindent and
$$
{\partial \over \partial a_{1}}  e^{\pm i\gamma /2} =
- e^{\pm i\gamma /2}  { \pm  i a_{2} \over  2\rho ^{2}} ,\qquad
{\partial  \over \partial a_{2}}  e^{\pm i\gamma /2}  =
+ e^{\pm i\gamma /2} {\pm i a_{1} \over  2\rho ^{2} }   ,
$$
$$
{\partial \over \partial a_{1}} \cos {\gamma \over 2}   =
+ {a_{2} \over 2\rho ^{2}} \sin { \gamma \over 2} \; ,\qquad
{\partial \over \partial a_{2}} \cos {\gamma \over 2} =
 - {a_{1} \over 2 \rho ^{2}}  \sin {\gamma \over 2} \;,
$$
$$
{\partial \over \partial a_{1}} \sin  {\gamma \over 2} =
- {a_{2} \over 2\rho ^{2}}  \cos {\gamma \over 2} \; ,\qquad
{\partial \over \partial a_{2}} \sin  {\gamma \over 2} =
+ {a_{1} \over 2\rho ^{2}}  \cos {\gamma \over 2} \; ;
$$

\noindent where  $\rho  = \sqrt{a^{2}_{1} + a^{2}_{2}}\; $.

Derivatives $\partial U^{1}/\partial a_{j}$ and
$\partial V^{1}/\partial a_{j}$  are
$$
{\partial U^{1} \over \partial a_{1}} =
\cos {\gamma \over 2}
 {a_{1} \over 2a \sqrt{ a + a_{3} }} +
         \sin {\gamma  \over 2}
{a_{2}\sqrt{ a + a_{3}} \over 2\rho ^{2}}  ,
$$
$$
{\partial U^{1} \over \partial a_{2}}  =
\;\cos{\gamma \over 2}
{a_{2} \over 2 a \sqrt{ a + a_{3}}}  -
\sin {\gamma \over 2}{a_{1} \sqrt{a + a_{3} } \over 2\rho ^{2}} ,
$$
$$
{\partial V^{1} \over \partial a_{1}} =
\; -\sin{\gamma \over 2}
{a_{1} \over 2a \sqrt{a + a_{3}}}  +
\cos{\gamma \over 2}{a_{2} \sqrt{ a + a_{3} } \over 2 \rho ^{2} } ,
$$
$$
{\partial V^{1} \over \partial a_{2}} =
\; -\sin {\gamma \over 2}
{a_{2} \over 2a \sqrt{a + a_{3} }}  -
\cos{\gamma \over 2}{a_{1}\sqrt{a + a_{3}} \over 2\rho ^{2}}  ;
$$

\noindent and  derivatives  $\partial U^{2}/\partial a_{j}$ ,
$\partial V^{2}/\partial a_{j}$  are
$$
{\partial  U^{2} \over \partial a_{1}} =
\;\cos{\gamma \over 2} {a_{1} \over 2a \sqrt{a - a_{3}}} +
\sin  {\gamma  \over 2} {a_{2} \sqrt{a - a_{3} } \over 2\rho ^{2}}  ,
$$
$$
{\partial U^{2} \over \partial a_{2}} =
\cos  {\gamma \over 2} {a_{2} \over 2a \sqrt{a - a_{3}}} -
\sin  {\gamma  \over 2} {a_{1} \sqrt{a - a_{3} } \over 2\rho ^{2}}  \; ,
$$
$$
{\partial V^{2} \over \partial a_{1}} =
\sin{\gamma \over 2} {a_{1} \over 2a \sqrt{a - a_{3}}} -
\cos{\gamma \over 2} {a_{2}\sqrt{ a - a_{3}} \over 2 \rho ^{2}}  ,
$$
$$
{\partial V^{2} \over \partial a_{2}} =
\sin {\gamma \over 2} {a_{2} \over 2a \sqrt{a - a_{3}}} +
\cos {\gamma \over 2} {a_{1} \sqrt{a - a_{3} } \over 2\rho ^{2}}  .
$$

\noindent
With the use of  these equations , we arrive at modified Cauchy-Riemann relations
\begin{eqnarray}
{\partial U^{1} \over \partial a_{1}} -
{\partial V^{1} \over \partial a_{2}}
= {1 \over 2}  ( a_{1} \cos  {\gamma  \over 2} +
                 a_{2} \sin  {\gamma  \over 2} )
\left [{1\over a \sqrt{a+a_{3}}} + {\sqrt{a+a_{3}} \over \rho ^{2}}\right ] ,
\nonumber
\\
{\partial  U^{1} \over \partial a_{2}} +
{\partial  V^{1} \over \partial a_{1}}
={1  \over  2}  ( a_{2} \cos {\gamma \over 2} -
                   a_{1} \sin {\gamma \over  2} )
\left [{1 \over a\sqrt{a + a_{3}}} + {\sqrt{a + a_{3}}\over \rho^{2}}\right ] ,
\nonumber
\\
{\partial U^{2} \over \partial a_{1}} -
{\partial V^{2} \over \partial a_{2}}
 ={1 \over 2} ( a_{1} \cos  {\gamma  \over 2}  -
                 a_{2} \sin  {\gamma  \over 2} )
\left [{1 \over a \sqrt{a - a_{3}}} - {\sqrt{a - a_{3}}\over \rho^{2}}\right ] ,
\nonumber
\\
{\partial U^{2} \over \partial a_{2}} +
{\partial V^{2} \over \partial a_{1}}
 = {1 \over 2}  ( a_{2} \cos  {\gamma  \over 2} +
                 a_{1} \sin  {\gamma  \over 2} )
\left [{1\over a \sqrt{a - a_{3}}} - {\sqrt{a -
 a_{3}} \over \rho^{2}}\right ]  .
\nonumber
\end{eqnarray}

\noindent
If  $a_{3}= 0$, we get
\begin{eqnarray}
{\partial U^{1} \over \partial a_{1}}  -
{\partial V^{1} \over \partial a_{2}}  =
{1 \over \sqrt{ \rho }} \; \cos {\gamma \over 2}  , \qquad
\nonumber
\\
{\partial U^{1} \over \partial a_{2}}  +
{\partial V^{1} \over \partial a_{1}}  =
{1 \over \sqrt{ \rho }} \; \sin {\gamma  \over 2} , \qquad
\nonumber
\\
{\partial U^{2} \over \partial a_{1}} -
{\partial V^{2} \over \partial a_{2}}  = 0    , \;\;\;
{\partial U^{2} \over \partial a_{2}} +
{\partial V^{2} \over \partial a_{1}}  = 0
\nonumber
\end{eqnarray}

\noindent which is quite understandable if we take into account
the  form of spinor $\xi $ at $a_{3} = 0$
\begin{eqnarray}
\xi ^{+\cap -}  =
                                \left ( \begin{array}{c}
( \sqrt{a_{1} +i a_{2}} \; )^{*} \\
\sqrt{a_{1} +i a_{2}}           \end{array} \right )  .
\label{4.9}
\end{eqnarray}

\noindent
At  $\rho\; \rightarrow \;\infty $
C-R condition will  hold.

Special note should be given to
behavior  of the spinor field $\xi ^{i}$ along half-plane
$\{ a_{1} \ge  0 , a_{2} = 0 \} ^{a_{3}}$. Here spinor $\xi$
is  not  a single-valued  function of  spatial points of the vector space $Pi_{3}$
because its values depend on direction from which one approaches the points.

\section{Calculating  $\nabla \xi$ and $\nabla_{\vec{n}} \; \xi$}

\hspace{7mm}
Spatial spinor field $\xi ^{a_{3}} (a_{1} + i a_{2})$
is not differentiable in the C-R sense.
However, some continuity property  of the spinor field yet exists.
With this in mind, let us calculate 2-gradient of
$\xi (a_{j})$:
\begin{eqnarray}
\nabla \xi =  ( {\partial \over \partial a_{1}} \;\xi\; , \;\;
{\partial \over \partial a_{2}} \; \xi )  , \;
\xi  = \xi ^{a_{3}} (a_{1}, \; a_{2}) \; .
\label{5.1}
\end{eqnarray}

\noindent  This quantity could serve
as characteristics of smoothness of spinor field
$(\xi ^{1}, \xi ^{2})$. With the use  of formulas
from previous Section one readily gets
\begin{eqnarray}
{\partial  \over \partial a_{1}} \; \xi ^{1}= {1 \over 2}
\xi ^{1}\;  ( {a_{1} \over a(a + a_{3})} +
i \; {a_{2} \over \rho ^{2}}  ) \; ,  \;\;
\nonumber
\\
{\partial  \over \partial a_{2}} \; \xi ^{1} = {1 \over 2}
\xi ^{1}  ( {a_{2} \over a(a + a_{3})} -
i \; {a_{1} \over \rho ^{2}}  )\;  , \;\;
\label{5.2a}
\\
{\partial \over \partial a_{1}} \; \xi ^{2} = {1 \over 2}
\xi^{2}  ( {a_{1} \over a(a + a_{3})} -
 i \; {a_{2} \over \rho ^{2}} ) \; , \;\;
\nonumber
\\
{\partial  \over \partial a_{2}} \; \xi ^{2} = {1 \over 2}
\xi ^{2}  ( {a_{2} \over a(a + a_{3})} -
i\; {a_{1} \over \rho^{2}}  )\; . \;\;\;
\label{5.2b}
\end{eqnarray}

\noindent
The form of these equations  will look  shorter if
one  uses gradient along directions
$\nabla _{\vec{n}}\; \xi = ( \vec{n} \; \nabla \xi )$ in the vicinity of every point.
From  (\ref{5.2a}) and (\ref{5.2b})  it follows
\begin{eqnarray}
\nabla_{\vec{n}} \; \xi ^{1} = {1 \over 2}\;
\left [\;{ (\vec{n} \; \vec{a}) \over a (a + a_{3})}\;+\;
i \; {\vec{n} \times  \vec{a} \over \rho ^{2}}\; \right ]
\; \xi^{1}\; , \;\;
\nonumber
\\
\nabla _{\vec{n}} \;\xi ^{2} =  {1 \over 2} \;
\left [\; {(\vec{n} \;\vec{a}) \over a (a - a_{3})}\;-\;
i\; {\vec{n} \times \vec{a}  \over \rho ^{2}}\; \right ] \xi ^{2}\;   ,\;\;\;
\label{5.3}
\end{eqnarray}

\noindent
where
$$
(\vec{n}\; \vec{a} ) =  n_{1} a_{1} + n_{2} a_{2}  , \;\;
\vec{n} \times \vec{a} =  n_{1} a_{2}  - n_{2} a_{1} .
$$
For  every vector  $\vec{a} = (a_{1}, a_{2})$
one  can consider  two directions $\vec{n}$,
parallel end orthogonal to it.
If $\vec{n} = \vec{n}_{\parallel}$, then $(\vec{n}\; \vec{a} ) = 0$ and
\begin{eqnarray}
\nabla _{\parallel} \xi ^{1} = {1 \over 2} \;
{ ( \vec{n} \; \vec{a} )  \over a( a + a_{3})}  \; \xi \; ,\qquad
\nabla _{\parallel} \xi ^{2} = {1 \over 2} \;
{(\vec{n} \; \vec{a}) \over a( a - a_{3})}  \;\xi \;  .
\label{5.4a}
\end{eqnarray}

\noindent
If  $\vec{n} = \vec{n}_{\perp}$ then
$(\vec{n} \; \vec{a} ) = 0$ and
\begin{eqnarray}
\nabla_{\perp} \xi^{1} = {i \over 2}
{\vec{n} \times \vec{a}  \over \rho ^{2} }  \xi ^{1}\; ,
\qquad
\nabla _{\perp} \xi ^{2} = - {i \over 2}
{\vec{n} \times \vec{a} \over \rho ^{2} }  \xi^{2}\;  .
\label{5.4b}
\end{eqnarray}

\noindent In other words, the  equations  have the structure
$$
 \nabla  _{\vec{n}} \xi =  \nabla _{\parallel} \xi \;+\; \nabla _{\perp} \xi \; .
$$

The relations  (\ref{5.3})   can be re-written in matrix  form
\begin{eqnarray}
\nabla_{\vec{n}} \; \xi  = A \; \xi \; .
\label{5.5}
\end{eqnarray}

Relation  (\ref{5.5})  can be considered   alternatively  as a master equation that
prescribes the explicit form of spinor $\xi (\vec{a})$  -- from which we had
started  in the beginning. This estimation of equation  (\ref{5.5})  seems
to be interesting and  possibly fruitful. As for now, it does not look simple or
fundamental indeed, however having been in their infancy it
does have  exiting mathematical
potential.

\section{ Spinor field $\eta$  peculiarities}

\hspace{7mm}
In this Section we are going to examine more closely singular properties of spinor
field $\xi ^{a_{3}} (a_{1}, a_{2})$.
At this, three cases, $\;a_{3} < 0 ,\; a_{3}  = 0 , \; a_{3} > 0 \;$
should  be considered
separately.

Evidently, there exist peculiarities  on the whole axis $(0,\; 0,\; a_{3})$  and along the  whole
half-plane $ ( \; a_{1} \ge 0 , \; a_{2} = 0 \; )^{a_{3}} $.
For every point of the axis,  instead of a single value,
spinor has a set of values (mute variable
$\Gamma$). At every point of the half-plane, instead of a single  value, spinor has two ones,
 different in sign  -- assuming the vector space  model
 is  investigated in terms  of spinor field.
Therefore, the quantity  $\nabla_{\vec{n}}\;\xi $ cannot be calculated
without  trouble in these peculiar sets $\{ \vec{a}^{\;0} \}$ --
where spinor $\xi$ losses single-valuedness.
As an alternative, for these points  there may be determined another characteristics
\begin{eqnarray}
\nabla _{\vec{n}}^{\vec{m}} \;\xi (\vec{a}^{\;0}) = \lim_{\epsilon \rightarrow 0} \nabla_{\vec{n}} \;
\xi (\vec{a}^{\; 0} \;+\; \epsilon \; \vec{m} )
 \; ,
\label{6.1}
\end{eqnarray}

\noindent that is one should find the  quantity
$\nabla_{\vec{n}} \xi $  in the vicinity of singular
point  $\vec{a}^{\;0}$ and  then  passes to
 the limit approaching to $\vec{a}^{\;0}$ along different ways.
In this line, let us consider the neighborhood of $(0,0)$ at $\Pi_{3}^{+}$:
$$
\vec{a} =  \vec{a}^{\;0} +\epsilon\; \vec{m}  \; , \;\;
\vec{a}^{\;0}= (0,\; 0)\; , \; \epsilon  \rightarrow  0 \; .
$$

\noindent
Taking $\epsilon $ as a small parameter we  get to $\tilde{\Pi}^{^{+}}_{0}: $
\begin{eqnarray}
e^{+i\gamma } \sim ( m_{1}+i m_{2}) = e^{+iM} \; ,\;
(a - a_{3}) \sim {\epsilon ^{2} \over 2 a_{3}} \; ,
\nonumber
\\
\xi ^{1} \sim \sqrt{2a_{3}} \; e^{-iM/2} \; ,\qquad
\xi ^{2} \sim {\epsilon  \over \sqrt{2a_{3}}} \;e^{+iM/2}\; .
\nonumber
\end{eqnarray}

\noindent Substituting these  into (\ref{6.1}), we  arrive at
\begin{eqnarray}
\tilde{\Pi}^{+}_{0} , \qquad
\nabla _{\vec{n}}^{\vec{m}} \;\xi^{1}(0,0) = {1 \over 2} \;
\sqrt{ 2a_{3}} \; e^{-iM/2}  \left[\;\epsilon\;{(\vec{n}\;
\vec{m} ) \over 2\; a^{2}_{3}} \;+\;i\;{(\vec{n} \times \vec{m})
\over \epsilon }\; \right ]\;  ,
\label{6.3a}
\\
\nabla _{\vec{n}}^{\vec{m}} \;\xi ^{2}(0,0) = {e^{+iM/2} \over
2 \sqrt{2a_{3}}}  \left [(\vec{n} \; \vec{m})-   i\;
 {\vec{n} \times \vec{m}  \over 2 } \right ]  .
\label{6.3b}
\end{eqnarray}

\noindent
Here the vector $\vec{m}$ cannot be taken as  $\vec{m}_{0} = (1,0)$  --  because,
if it is  so, the vector  $\vec{a} = (\vec{a}^{\;0} \;+\; \epsilon  \;
\vec{m}_{0})$  will get into a singular set
where $\nabla_{\vec{n}} \;\xi $ is not well
defined. Instead, one should analyze two limits only:
\begin{eqnarray}
\lim_{\vec{m}\; \rightarrow \; \vec{m}_{0}^{\;+}}
\nabla_{\vec{n}}^{\vec{m}} \; \xi ^{i}(0,0) = -
\lim_{\vec{m} \; \rightarrow \; \vec{m}_{0}^{-}}
 \nabla _{\vec{n}}^{\vec{m}}
\; \xi ^{i}(0,0)  \; .
\label{6.4}
\end{eqnarray}

\noindent Designation  $\vec{m} \rightarrow \vec{m}^{+}_{0}$
means that  $\vec{m}$  approaches to  $\vec{m}_{0}$  from up half-plane, whereas
$\vec{m} \rightarrow \vec{m}^{-}_{0}$ assumes that  $\vec{m}$
approaches to $\vec{m}_{0}$ from
lower half-plane.

In the same way, consideration of the neighborhood of $(0,0)$ in $\Pi_{3}^{-}$
leads to
\begin{eqnarray}
\tilde{\Pi}^{-}_{0}   : \;
\nabla _{\vec{n}}^{\vec{m}} \xi ^{1}(0,0)
 = { e^{-iM/2} \over \sqrt{-2a_{3}} }
\left [
 \vec{n}\; \vec{m} + {i \over 2}  \vec{n} \times
\vec{m} \right ] ,
\label{6.5a}
\end{eqnarray}
\begin{eqnarray}
\nabla_{\vec{n}}^{\vec{m}} \xi ^{2}(0,0)
 = {1 \over 2}  \sqrt{ -2a_{3}} e^{+iM/2} \;
\left [ \epsilon  { \vec{n} \;\vec{m} \over 2 a^{2}_{3}}
- i {\vec{n} \times \vec{m} \over \epsilon} \right ]    .
\label{6.5b}
\end{eqnarray}

As for the  point  $\tilde{\Pi}^{+\cap -}_{0}$ we will have
\begin{eqnarray}
\xi ^{1} \sim  \sqrt{\epsilon } \; e^{-iM/2} \; ,\;\; \xi ^{2}
\sim \sqrt{\epsilon } \; e^{+iM/2}
\label{6.6a}
\end{eqnarray}

\noindent and further
\begin{eqnarray}
\nabla _{\vec{n}}^{\vec{m}} \xi ^{+\cap -}_{0}
={ 1 \over 2 \sqrt{\epsilon}}  \left ( \begin{array}{c}
e^{-iM/2}   [ \vec{n}\; \vec{m} +
i \;\vec{n} \times \vec{m}  ] \\[2mm]
e^{+iM/2}  [\vec{n}\; \vec{m}  -
i \; \vec{n} \times \vec{m} ]
\noindent \end{array} \right )  .
\label{6.6b}
\end{eqnarray}

 In  a  sense, for every plane $(a_{2}, a_{2})^{a_{3}}$  its infinite boundary  is
peculiar as well ---  expression for $\nabla_{\vec{n}} \xi $  at the  line
$\{ \infty \;m_{1} , \infty\;  m_{2} \} ^{a_{3}}$ will be
($\; \Omega \; \rightarrow \; \infty$ )
\begin{eqnarray}
\nabla_{\vec{n}}^{\vec{m}}  \xi (\infty )
 = { 1 \over 2 \sqrt{\Omega }}  \left ( \begin{array}{c}
e^{-iM/2}[ \vec{n}\; \vec{m}+
i \; \vec{n} \times \vec{m} ] \\[2mm]
e^{+iM/2}[\vec{n} \; \vec{m} -
i \; \vec{n} \times \vec{m} ]
\end{array} \right )  .
\label{6.7}
\end{eqnarray}

Now, is is the point to examine spinor peculiarities  at the  half-plane
$\{ a_{1} > 0, a_{2} = 0 \}^{a_{3}}$:

\vspace{-7mm}

\unitlength=0.65 mm
\begin{picture}(160,70)(-60,0)
\special{em:linewidth 0.4pt}
\linethickness{0.4pt}

\put(+1,+30){\vector(+1,0){90}}  \put(+90,+25){$a_{1}$}
\put(+50,+10){\vector(0,+1){40}}  \put(+41,+47){$a_{2}$}
\put(+50,+30){\oval(20,20)} \put(+60,+25){\vector(0,+1){5}}
\put(+70,+35){$\gamma = 0$}
\put(+50,+29){\line(+1,0){35}} \put(+70,+20){$\gamma = 2 \pi$}
\end{picture}

\begin{center}
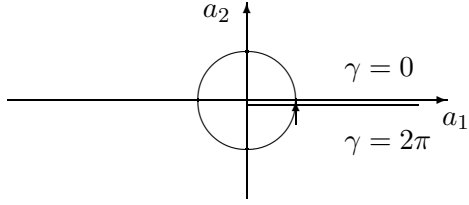
FIG. 7. Spinor peculiarities at
$ a_{2} = 0\;:\;\xi (\gamma = 0) = - \xi
(\gamma = 2 \pi ) $
\end{center}


\noindent
Here spinor field is double-valued.
To describe  that behavior let us act in  the way used above:
$$
\lim_{\epsilon \; \rightarrow \; 0} \; \nabla_{\vec{n}} \;
\xi (\vec{a}^{\; 0}\; +\; \epsilon \;\vec{m}) =
\nabla_{\vec{n}}^{\vec{m}} \; \xi(\vec{a}^{\;0})\; ,
$$
$$
\vec{a}^{\; 0} = ( a^{0}_{1} > 0 , \; a^{0}_{2} = 0 )\; , \;\;
\vec{m} \neq  \pm \; \vec{m}_{\; 0} = \pm ( 1, 0) \; .
$$

\noindent Taking into consideration
$$
 a_{1} \sim a^{0}_{1} + \epsilon \; m_{1} )\; , \qquad
a_{2} \sim \epsilon \;  m_{2}\; ,
$$
$$
\vec{n}\;  \vec{a} \sim  + n_{1}  a^{0}_{1} +
\epsilon \;  \vec{n} \; \vec{m}  \; ,\qquad
 \vec{n} \times  \vec{a} \sim - n_{2} a^{0}_{1} +
\epsilon\;  \vec{n}  \times  \vec{m} \;   ,
$$

\noindent and
$$
\lim_{\epsilon \rightarrow 0}  \xi ^{1}
(\vec{a}_{0} \; +\; \epsilon  \;
\vec{m}) = \sqrt{ a^{0} +  a^{0}_{3} } \; sgn \; ( m_{2})\; ,
$$
$$
 \lim_{\epsilon \rightarrow 0}\;
\xi ^{2}(\vec{a}_{0} \;+\; \epsilon \;
\vec{m}) =  \sqrt{a^{0} - a^{0}_{3} }\; sgn \; (m_{2})\; ,
$$

\noindent  we  easily obtain
\begin{eqnarray}
\nabla_{\vec{n}}^{\vec{m}}  \xi ^{1} = { a^{0}_{1} \over 2
\sqrt{a^{0} + a^{0}_{3}}} \; sgn \; (m_{2})
( { n_{1} \over a^{0}}  - i
 { n_{2} \over a^{0} - a^{0}_{3} } ) \;,
\nonumber
\\
\nabla _{\vec{n}}^{\vec{m}} \xi ^{2} =
{ a^{0}_{1} \over 2 \sqrt{a^{0} - a^{0}_{3}}} \;
 sgn\; (m_{2}) ( { n_{1} \over
a^{0}} + i { n_{2} \over a^{0} + a^{0}_{3} } ) \; .
\nonumber
\end{eqnarray}

\noindent These relations can be  accompanied  be the   diagram

\vspace{+3mm}

\unitlength=0.4 mm
\begin{picture}(160,100)(-110,-50)
\special{em:linewidth 0.4pt}
\linethickness{0.4pt}

\put(-40,0){\vector(+1,0){130}}
\put(+92,-5){$a_{1}$}
\put(0,-40){\vector(0,+1){80}}
\put(-10,+40){$a_{2}$}

\put(+40,0){\circle*{2}}
\put(+40,0){\oval(20,20)}
\put(+40,+25){\vector(0,-1){23}}
\put(+40,-25){\vector(0,+1){23}}
\put(+60,+20){\vector(-1,-1){18}}
\put(+20,-20){\vector(+1,+1){18}}
\put(+20,+20){\vector(+1,-1){18}}
\put(+60,-20){\vector(-1,+1){18}}

\put(+35,+30){$\{\vec{m}^{(+)} \}$}
\put(+35,-35){$\{\vec{m}^{(-)} \}$}
\end{picture}

\begin{center}
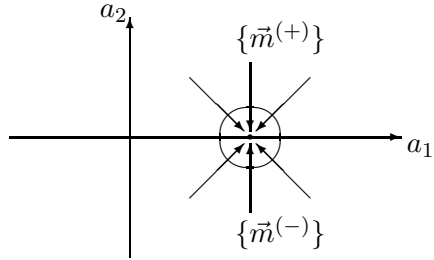
FIG. 8. Spinor peculiarities and $\pi$-vicinities
\end{center}

\vspace{3mm}

\noindent That is one may isolate two  angular   $\pi$-vicinities   near
the  point $\vec{a}_{0}$  ---  within each of them there is no  dependence
on $\vec{m}$, but
\begin{eqnarray}
\nabla_{\vec{n}}^{\vec{m}^{(-)}} \xi = -
\nabla_{\vec{n}}^{\vec{m}^{(+)}} \xi \;.
\label{6.8b}
\end{eqnarray}

Now, one can make some general remarks on the  mapping
$
\Pi_{3} \Longrightarrow \xi , \qquad
\tilde{\Pi}_{3} \Longrightarrow \xi
$
over the vector  $\Pi_{3}$  and  spinor  $\tilde{\Pi}_{3}$ space models.
The mapping $\Pi_{3} \Longrightarrow \xi$  may be illustrated by the  diagrams

 \vspace{+2mm}

\unitlength=0.53 mm
\begin{picture}(160,100)(-50,0)
\special{em:linewidth 0.4pt}
\linethickness{0.4pt}

\put(+10,+50){\vector(+1,0){50}}  \put(+120,+55){\vector(0,+1){40}}
\put(+65,+45){$a_{1}$}
\put(+30,+20){\vector(0,+1){60}}  \put(+23,+80){$a_{2}$}
\put(+30,+48){\line(+1,0){28}}
\put(+45,+55){$\gamma = 0$}       \put(+45,+40){$\gamma = 2\pi $ }
\put(+30,+50){\oval(20,20)}
\put(+40,+45){\vector(0,+1){5}}   \put(+40,+50){\circle*{1}}

\put(+100,+90){$\xi ^{1}$}
\put(+90,+75){\vector(+1,0){60}}
\put(+100,+78){$\gamma = 2 \pi$}   \put(+135,+78){$\gamma = 0$}
\put(+120,+75){\oval(10,40)[b]}    \put(+115,+70){\vector(0,+1){5}}

\put(+90,+55){\line(0,+1){20}}
\put(+95,+55){\line(0,+1){20}}
\put(+100,+55){\line(0,+1){20}}
\put(+105,+55){\line(0,+1){20}}
\put(+110,+55){\line(0,+1){20}}
\put(+130,+55){\line(0,+1){20}}
\put(+135,+55){\line(0,+1){20}}
\put(+140,+55){\line(0,+1){20}}
\put(+145,+55){\line(0,+1){20}}
\put(+150,+55){\line(0,+1){20}}

\put(+90,+25){\vector(+1,0){60}}  \put(+120,+5){\vector(0,+1){40}}
\put(+100,+18){$\gamma = 2 \pi$}  \put(+135,+18){$\gamma = 0$}
\put(+120,+25){\oval(10,40)[t]}   \put(+115,+30){\vector(0,-1){5}}
\put(+90,+45){\line(0,-1){20}}
\put(+95,+45){\line(0,-1){20}}
\put(+100,+45){\line(0,-1){20}}
\put(+105,+45){\line(0,-1){20}}
\put(+110,+45){\line(0,-1){20}}
\put(+130,+45){\line(0,-1){20}}
\put(+135,+45){\line(0,-1){20}}
\put(+140,+45){\line(0,-1){20}}
\put(+145,+45){\line(0,-1){20}}
\put(+150,+45){\line(0,-1){20}}

\put(+100,+10){$\xi ^{2}$}
\end{picture}

\begin{center}
FIG. 9. Spinor discontinuity
\end{center}

\noindent that is the whole real plane  $(a_{1}, a_{2})$ maps into
a couple of complex half-planes $\xi ^{1}$  and $\xi ^{2}$,
differently oriented. For these maps the existence of discontinuity
along a  positive half-axis
$$
\Pi_{3} , \; a_{2}= 0 ,  a_{1}
\geq 0 \; : \qquad  \xi (\gamma =  0) = - \xi (\gamma = 2\pi )
$$

\noindent  is inevitable. In contrast to this, the mapping
$\tilde{\Pi}_{3} \Longrightarrow \xi$ looks  more smooth:
$$
\tilde{\Pi}_{3},  a_{2}= 0,  a_{1} \geq 0  :  \;   \xi (\gamma =  -2\pi ) = +
 \xi (\gamma = + 2\pi ) .
$$

In other words, changing vector model into spinor one
 may be considered as a  way to ensure
continuity property of  spinor  field $\xi$
in  maximally large domain. In this  context, the  use of
2-sheeted planes instead of 1-sheeted planes appears to be natural and
intelligible operation.

Initial vector space $\Pi _{3}$  could  be thought of as a  collection
of all 1-sheeted  $a_{3}$-planes:
$$
\Pi_{3} = \sum_{a_{3}\in (-\infty ,
+\infty )}(a_{1},a_{2})^{a_{3}} \; ,
$$

\noindent  instead an extended space $\tilde{\Pi}_{3}$,
one may imagine spinor one as a collection of all 2-sheeted  $a_{3}$-planes:
$$
\tilde{\Pi}_{3} = \sum_{a_{3}\in (-\infty ,+\infty )}^{\delta = 1,2}
(a_{1}, a_{2})^{a_{3}}   \; .
$$

Any  2-sheeted plane differs in topological sense from 1-sheeted -- now
neighborhood of  a zero point $(0,0)$ is  not Euclidean. Therefore, extended space
$\tilde{\Pi}_{3}$ will be  non-Euclidean as well.
The concept of nearness in such a model should  take special attention to:
nearness in Euclidean sense  $\Pi_{3}$ is  not the same as  the  nearness  for
extended  model $\tilde{\Pi}_{3}$.
Indeed,  two points can be near to each other only
if they both belong to the same sheet or if they
approach to  a sewing domain. For example,
the following  points
$$
\{ a^{\delta =1} _{1}, a^{\delta = 1}_{2}  \}_{a'_{3}} , \;
\{ a^{\delta =1} _{1} , a^{\delta = 1}_{2}  \}_{a''_{3}}  ,
\;\;  \mbox{if} \; \; (a''_{3} - a'_{3} ) \rightarrow  0
$$

\noindent are neighboring ones; analogously close will be the points
(if $ \; (a''_{3} - a' _{3} ) \; \rightarrow \; 0\;$)
$$
 \; \{ \; a^{\delta =2}_{1} , a^{\delta = 2}_{2} \; \}_{a'_{3}} \; , \qquad
\{\; a^{\delta =2}_{1} , a^{\delta = 2}_{2} \; \}_{3''_{3}}\; .
$$

\noindent
However, two points
$
\{ \; a^{\delta =1}_{1} , a^{\delta = 1}_{2} \; \}_{a'_{3}}\;$ and  $\;
\{\; a^{\delta =2}_{1} , a^{\delta = 2}_{2} \; \}_{a'_{3}} \;$
will be quite distant from each other if  they  do not belong   to a sewing domain.

\vspace{5mm}
In a precise  form,   changing  space  $\Pi_{3}$ into
extended space $\tilde{\Pi}_{3}$ results  in

\vspace{5mm}

\noindent
\underline{for model  $\Pi_{3}$}
\vspace{5mm}

\noindent
1)  spinor  $\xi (\vec{a})$  is exponentially discontinuous   at the points
 $(0,0)_{a_{3}}$ and $(\pm)\; $-valued along half-plane $(0,a_{2} = 0)^{a_{3}}$;

\noindent
2)  spinor $\xi = \xi(\vec{a},\vec{m})$ has discontinuity
on  a unique direction near to $(0, 0)_{a_{3}}$  and
on two direction near the  half-plane  $(0,a_{2} = 0)^{a_{3}}$.

\vspace{5mm}

\noindent
\underline{for model  $\tilde{\Pi}_{3}$}
\vspace{5mm}

\noindent
1) spinor $\xi (\vec{a}^{\;\delta =1,2} )$ is exponentially discontinued
at the points $(0,0)_{a_{3}}$; any points  of
$(\pm)\; $-valued discontinuity does not exist;

\noindent
2)  spinor $\xi (\vec{a}^{\;\delta =1,2}, \vec{m})$  is  continuous everywhere.

\vspace{5mm}

So, the change of a space model $\Pi_{3}$
substantially alters underlined spinor field's continuity properties.
In the next sections, in the same line, we are going to examine spinor geometry
of properly vector  space $E_{3}$. It  seems  important,  in a  parallel way
to have both spinor models, resulting respectively from different $P$-orientations
of an initial space. The main idea is to make explicit manifestations
of geometrical difference  of pseudo and  properly vector space models
apparent as  much as  possible.

\section{Spinor  $\eta ^{b_{3}} (b_{1} + i b_{2})$ and   analyticity }

\hspace{7mm}
Let us consider spinor components $\eta (b_{1}, b_{2}, b_{3})$
as complex-valued functions of  $z = ( b_{1} + i b_{2})$ and
parameter  $b_{3}$   (let $\sigma = \pm 1$):
\begin{eqnarray}
\eta^{(\sigma )}(b_{j}) = \left ( \begin{array}{r}
\eta ^{1(\sigma )}(b_{3},  b_{1}  + i b_{2} ) \\[2mm]
\eta ^{2}(b_{3}, b_{1} +   i b_{2} )  \end{array} \right ) \; .
\label{7.1}
\end{eqnarray}

\noindent The notation will be  used
\begin{eqnarray}
\eta ^{1(\sigma )} =
\sigma \; g^{-} (\cos {\gamma \over 2} - i\sin {\gamma\over 2})
=  U^{1(\sigma )} + i V^{1(\sigma )}  \; ,
\nonumber
\\
\eta ^{2} = g^{+} (\cos{\gamma  \over 2} + i \sin{\gamma \over 2} )
=  U^{2} + i V^{2}  \; , \qquad
\nonumber
\\
g^{\pm} = \sqrt{b \pm (b^{2}_{1} + b^{2}_{2})} \; , \qquad \qquad\qquad
\nonumber
\\
e^{i\gamma /2} = \sqrt{{b_{1} + i b_{2} \over
(b^{2}_{1} + b^{2}_{2})^{1/2} }} =  \cos {\gamma \over 2} +
i \sin  {\gamma  \over 2}  \;  . \qquad
\nonumber
\end{eqnarray}

\noindent  Derivatives will be  needed
\begin{eqnarray}
{\partial \over \partial  b_{1}} g^{\pm} = \pm {b_{1} \over b}\;
{\sqrt{b \pm (b^{2}_{1}+ b^{2}_{2})^1/2 } \over 2
\sqrt{b^{2}_{1} + b^{2}_{2} }}  \; ,\;\;
{\partial  \over \partial b_{2}} g^{\pm} = \pm {b_{2} \over b}\;
{\sqrt{b \pm (b^{2}_{1} + b^{2}_{2})^1/2}  \over 2
\sqrt{b^{2}_{1} + b^{2}_{2}}}  \; ,
\nonumber
\\
{\partial  \over \partial  b_{1}} e^{\pm
i\gamma /2} = - e^{\pm i\gamma /2}\; {\pm i b_{2} \over 2
\rho ^{2}} \; , \qquad
{\partial \over \partial  b_{2}} e^{\pm i\gamma /2} = -
e^{\pm i\gamma /2}\; {\pm i b_{1} \over 2
\rho ^{2}} \; , \qquad
\nonumber
\\
{\partial  \over \partial b_{2}} \;\cos {\gamma \over 2} = -
{b_{1} \over 2 \rho ^{2}} \; \sin {\gamma  \over 2} \; , \qquad
{\partial \over \partial b_{2}} \; \sin {\gamma  \over 2} = +
{b_{1} \over 2\rho ^{2}} \; \cos {\gamma \over 2} \; , \qquad
\nonumber
\end{eqnarray}
\begin{eqnarray}
{\partial \over \partial b_{1}}  U^{1(\sigma )} =
\sigma {\sqrt{b - \rho } \over 2 \rho } \;
\left [  - { b_{1} \over b} \cos {\gamma \over 2}  + {b_{2}\over \rho }
\sin  {\gamma  \over 2}  \right ]   ,
\nonumber
\\
{\partial \over \partial b_{2}} U^{1(\sigma )} =
\sigma {\sqrt{b - \rho } \over 2 \rho }
\left [ -{ b_{2}\over b}\cos {\gamma \over 2} -
{b_{1}\over \rho }  \sin  {\gamma  \over 2}  \right ]  ,
\nonumber
\\
{\partial \over \partial b_{1}} V^{1(\sigma )} =
\sigma {\sqrt{b - \rho } \over 2 \rho }
\left [-{b_{1}\over b} \sin {\gamma \over 2} + {b_{2}\over \rho }
\cos {\gamma \over 2} \right ]  ,
\nonumber
\\
{\partial \over \partial b_{2}}  V^{1(\sigma )} =
\sigma {\sqrt{b - \rho } \over 2 \rho }
\left [+{b_{2} \over b} \sin {\gamma \over 2} - {b_{1}\over \rho }
\cos {\gamma  \over 2} \right ]  ;
\nonumber
\end{eqnarray}
\begin{eqnarray}
{\partial  \over \partial  b_{1}}  U^{2} =
{\sqrt{b + \rho } \over 2 \rho }
\left [  +{b_{1} \over b}  \cos {\gamma \over 2}  +
{b_{2} \over \rho } \sin  {\gamma  \over 2}  \right ]  ,
\nonumber
\\
{\partial \over \partial b_{2}} U^{2} =
{\sqrt{b + \rho } \over 2 \rho }
\left [+{ b_{2}\over b} \cos {\gamma \over 2} -{b_{1} \over \rho }
 \sin {\gamma  \over 2} \right ]  ,
\nonumber
\\
{\partial  \over \partial b_{1}}  V^{2} =
{\sqrt{b + \rho } \over 2 \rho }
\left [ +{b_{1} \over b} \sin {\gamma \over 2} -  {b_{2}\over \rho }
\cos {\gamma  \over 2}  \right ]  ,
\nonumber
\\
{\partial  \over \partial  b_{2}}
V^{2} = {\sqrt{b + \rho } \over 2 \rho }
\left [  +{b_{2}\over b}  \sin {\gamma \over 2} +  {b_{1}\over \rho }
 \cos {\gamma  \over 2}  \right ]  .
\nonumber
\end{eqnarray}

\noindent Again,  we find the modified Cauchy-Riemann relations
\begin{eqnarray}
{\partial \over \partial b_{1}} U^{1(\sigma )}
- {\partial  \over \partial b_{2}} V^{1(\sigma )}  =
\sigma {\sqrt{ b - \rho } \over 2 \rho } \;
({1 \over \rho } - {1 \over b})
\left [ b_{1} \cos {\gamma  \over 2} +
b_{2} \sin {\gamma  \over 2} \right ]  ,
\nonumber
\\
{\partial \over \partial b_{2}} U^{1(\sigma )} +
{\partial  \over \partial b_{1}} V^{1(\sigma )}  =
\sigma {\sqrt{ b - \rho } \over 2 \rho } ({1\over \rho } -
{1 \over b}) \left [  b_{1} \cos {\gamma  \over 2} - b_{2} \sin
{\gamma \over 2}  \right ]  \; ,
\label{7.5a}
\end{eqnarray}
\begin{eqnarray}
{\partial \over \partial b_{1}} U^{2} -
{\partial  \over \partial  b_{2}}  V^{2}  =
{\sqrt{ b + \rho } \over 2 \rho } ({1 \over \rho } - {1 \over b})
\left [-b_{1} \cos {\gamma  \over 2}   +
 b_{2} \sin {\gamma \over 2} \right ]  \; ,
\nonumber
\\
{\partial \over \partial b_{2}}  U^{2} +
{\partial \over \partial b_{1}}  V^{2}  =
{\sqrt{ b + \rho } \over  2 \rho  }
({1 \over \rho } - {1 \over b})
\left [ -b_{1} \cos {\gamma \over 2} - b_{2} \sin {\gamma \over 2}
\right ]  \; .
\label{7.5b}
\end{eqnarray}

\noindent  When  $b_{3}= 0$, from (\ref{7.5a}) and (\ref{7.5b}) it follows
$$
{\partial  \over \partial b_{1}} U^{1(\sigma )} -
{\partial  \over \partial b_{2}} V^{1(\sigma )} = 0  ,\qquad
{\partial  \over \partial b_{2}} U^{1(\sigma )} +
{\partial  \over \partial b_{1}} V^{1(\sigma )} = 0  ,
$$
$$
{\partial \over \partial b_{1}} U^{2} -
{\partial \over \partial b_{2}} V^{2} = 0  , \qquad
{\partial \over \partial b_{2}} U^{2} -
{\partial \over \partial b_{1}} V^{2} = 0  ,
$$

\noindent that is C-R relations hold. It is  consistent  with
the  form of spinor $\eta$ at  $b_{3}= 0$:
$$
\eta ^{+\cap -}  = \sqrt{2 \rho }
\left ( \begin{array}{c}
0 \\ e^{+i\gamma /2}  \end{array} \right ) =
\sqrt{2} \left ( \begin{array}{c}
0  \\ \sqrt{b_{1} + i b_{2}} \end{array} \right ) \; .
$$

\section{Spinor $\eta$  continuity properties }

\hspace{7mm}
The 2-gradient of spinor  field $\eta$ will be
(symbol  $\sigma$  at  $\eta ^{1}$ is omitted):
\begin{eqnarray}
{\partial \over \partial b_{1} } \eta ^{1} =
\eta ^{1}  {1 \over 2 \rho }
\left ( - {b_{1} \over b}  +  i  {b_{2} \over b} \right )\; ,\qquad
{\partial \over \partial  b_{2} }  \eta ^{1} = \eta ^{1}
{1 \over 2 \rho }  \left (  -{b_{2} \over b}  - i {b_{1} \over b}
\right ) \; ,
\nonumber
\\
{\partial  \over \partial  b_{1} } \eta ^{2} =
\eta ^{2}  {1 \over 2 \rho } \left ( + {b_{1} \over b} -
i {b_{2} \over b} \right )\; ,\qquad
{\partial  \over \partial  b_{2} }  \eta ^{2} = \eta ^{2}
{1 \over 2 \rho }  \left (  + {b_{2} \over b} +
i {b_{1} \over b} \right )  .
\label{8.1}
\end{eqnarray}

\noindent From  (\ref{8.1}) it follows
\begin{eqnarray}
\nabla _{\vec{n}}  \eta ^{1} = \eta ^{1} {1 \over 2 \rho }
\left [ - {1 \over b}  (\vec{n} \;\vec{b})  +
{i \over \rho }   (\vec{n} \times \vec{b}) \right ] \; ,\qquad
\nabla _{\vec{n}}  \eta ^{2} = \eta ^{2} {1 \over 2 \rho}
\left [ + {1 \over b}  (\vec{n} \;\vec{b})  -
{i \over \rho}  (\vec{n} \times \vec{b}) \right ] \; .
\label{8.2}
\end{eqnarray}

\noindent Here again (see in Section 6) one  can see two terms:
\begin{eqnarray}
 \nabla _{\vec{n}}  \eta  = ( \nabla _{\perp} \eta +
\nabla _{\parallel}  \eta )  .
\nonumber
\end{eqnarray}

\noindent In the case  $b_{3}=0$ relations (\ref{8.2}) look  much simpler
\begin{eqnarray}
\eta ^{1}_{b_{3} = 0} = 0  , \qquad
\nabla_{\vec{n}}  \eta ^{2}_{b_{3} = 0} =\eta ^{2}_{b_{3} = 0}  {1 \over 2\rho ^{2} } \;
[\; \vec{n} \;\vec{b} -  i \;\vec{n} \times \vec{b} \; ] .
\label{8.3}
\end{eqnarray}

\section{Peculiarities of field
$\eta ^{b_{3}} (b_{1} + i b_{2})$}

\hspace{7mm}
Consideration of the problem will be performed  in the  manner used in
 Section 6.
In  the  neighborhood of $(0,0)^{b_{3}} \in \tilde{E}^{+}_{3}$ there is
\begin{eqnarray}
\vec{b} \sim ( \epsilon \; m_{1} , \; \epsilon \; m_{2}, \; b_{3}) \; ,
\;\;\;  b_{3} > 0 \; , \qquad
\nonumber
\\[1mm]
\eta ^{1} \sim + \sqrt{b_{3}}\; e^{-iM/2} \; , \;
\eta ^{2} \sim + \sqrt{b_{3}}\; e^{+iM/2} \; ,
\nonumber
\\
\nabla_{\vec{n}}^{\vec{m}}  \eta ^{1} \sim
{\sqrt{b_{3}} e^{-iM/2} \over 2} \left [ -
{(\vec{n} \vec{m}) \over b_{3}} + i
{(\vec{n} \times \vec{m})  \over \epsilon} \right ]
 ,
\nonumber
\\
\nabla_{\vec{n}}^{\vec{m}}  \eta ^{2} \sim {\sqrt{b_{3}}
e^{+iM/2} \over 2}  \left [ + {(\vec{n}  \vec{m}) \over b_{3}} -
{(\vec{n} \times \vec{m})  \over \epsilon} \right ],
\label{9.1b}
\end{eqnarray}

\noindent here $ \vec{m} \neq ( +1, 0, 0 )$.
Analogously, for
$(0,0)^{b_{3 }} \in \tilde{E}^{-}_{3}$ there is
\begin{eqnarray}
\vec{b} \sim ( \epsilon\; m_{1} , \epsilon\; m_{2}, b_{3} ) \;,
\;\;  b_{3} < 0  , \qquad
\nonumber
\\[1mm]
\eta ^{1} \sim - \sqrt{-b_{3}}  e^{-iM/2}  , \;
\eta ^{2} \sim + \sqrt{-b_{3}}  e^{+iM/2}  ,
\nonumber
\end{eqnarray}
\begin{eqnarray}
\nabla _{\vec{n}}^{\vec{m}}  \eta ^{1} \sim
{\sqrt{-b_{3}}  e^{-iM/2} \over 2} \left [
+ {(\vec{n} \vec{m}) \over \mid b_{3} \mid } -
i {(\vec{n} \times \vec{m})  \over \epsilon} \right ]\; ,
\nonumber
\\
\nabla _{\vec{n}}^{\vec{m}} \eta ^{2} \sim
{\sqrt{-b_{3}} e^{+iM/2} \over 2} \left [
+ {(\vec{n} \vec{m}) \over \mid b_{3} \mid } -
{(\vec{n} \times \vec{m}) \over \epsilon}  \right ]  .
\nonumber
\label{9.2b}
\end{eqnarray}

\noindent Near the  points $E^{+\cap -} _{0}$,
when $ \vec{b} \sim ( \epsilon \; m_{1} , \; \epsilon \; m_{2},\; 0 )$,
 we have
\begin{eqnarray}
\eta ^{1} = 0 \; ,\;  \eta ^{2} = \sqrt{2 \epsilon }
e^{+iM/2} \; , \; \nabla _{\vec{n}}^{\vec{m}}  \eta ^{1} = 0 \; ,
\label{9.3a}
\\
\nabla _{\vec{n}}^{\vec{m}}\; \eta ^{2} =
{e ^{+iM/2} \over 2 \epsilon }   \left [\;
\vec{n}\;\vec{m}  -  i \; \vec{n} \times \vec{m} \; \right ] \; .
\nonumber
\end{eqnarray}

\noindent
For half-axis  $\{ b^{0}_{1} > 0 ,\; b^{0}_{2}= 0 \}$ we  will have (the notation
$b^{0} = \sqrt{(b^{0}_{1})^{2} + (b^{0}_{3})^{2}}$ is used)
\begin{eqnarray}
\nabla _{\vec{n}}^{\vec{m}}  \eta^{1(\sigma )} =
\sigma \;  {\sqrt{b^{0} - b^{0}_{1} } \over 2}  \left [
- {n_{1} \over b_{0}}  -  i  {n_{2} \over b^{0}_{1} }
 \right ] \;\mbox{sgn} \; (m_{2})     ,
\nonumber
\\
\nabla _{\vec{n}}^{\vec{m}} \eta ^{2 } =
\sigma \; {\sqrt{b^{0} + b^{0}_{1} } \over 2} \left [
+ {n_{1} \over b_{0}}\; + \;{n_{2} \over b^{0}_{1}}
\right ]  \;\mbox{sgn} \; (m_{2}) \;  . \;\;\;
\nonumber
\label{9.3b}
\end{eqnarray}

Everything said in the  end of Section 5 on the  pseudo vector  model
is  applied here too;  it is unnecessary  to repeat the same  else one  time.

\section{ Comparing models $\xi$  and $\eta$}

\hspace{7mm} Now we are going to describe some qualitative
distinctions between spinor models $\xi$ and $\eta$. Two models of
spinors spaces  with respect to $P$-orientation are grounded  on
different mappings  $\xi$   and  $\eta $  defined over  the same
extended domain $\tilde{G}(y_{i})$.  The natural question is:  how
are  these two  maps  connected to each others. An answer  can be
found on comparing the formulas  for $\xi $  and $\eta $.  An
answer can be straightforwardly found. Indeed, taking into account
identities
\begin{eqnarray}
{1 \over \sqrt{2}} (\sqrt{x + x_{3}} + \sqrt{x -
x_{3}}) = + \sqrt{ x + \rho }  , \; x_{3} > 0  ,
\nonumber
\\
 {1
\over \sqrt{2}} (\sqrt{x + x_{3}} - \sqrt{x - x_{3}}) = -
\sqrt{x - \rho } , \; x_{3} < 0  ,
\nonumber
\\
 {1 \over \sqrt{2}}
(\sqrt{x + x_{3}} + \sqrt{x - x_{3}}) = + \sqrt{ x + \rho } , \qquad
\nonumber
\end{eqnarray}

\noindent one can straightforwardly arrive at
$$
\eta _{1} = { \xi _{1} - \xi _{2}^{*} \over \sqrt{2}} \;, \qquad
\eta_{2}  = {\xi _{1}^{*} + \xi _{2} \over \sqrt{2}}
$$

\noindent or in more short form
\begin{eqnarray}
\eta = {1 \over \sqrt{2}} ( \xi - i \; \sigma ^{2} \xi ^{*} )\; .
\label{10.1a}
\end{eqnarray}

\noindent Inverse to  (\ref{10.1a}) looks as
$$
\xi _{1} = { \eta _{1} + \eta _{2}^{*} \over \sqrt{2}} \; ,\qquad
\xi _{2} = { \eta _{2} - \eta _{1}^{*} \over \sqrt{2}} \; ,
$$

\noindent
or
\begin{eqnarray}
\xi = {1 \over \sqrt{2}} \; ( \eta \;-\; i \;\sigma ^{2} \eta ^{*} )
\; .
\label{10.1b}
\end{eqnarray}

In connection with eqs. (\ref{10.1a}) and (\ref{10.1b})
there are two points to which special attention
 must be  given:

1) complex conjugation enters them explicitly which
correlates with the  change in orientation properties of the  models;

2) spinors $\xi $   and  $i \sigma ^{2} \xi ^{*}$ (as well as  $\eta$
and $i \sigma ^{2} \eta ^{*}$) provide us with non-equivalent representations
of the extended   unitary group $\tilde{SU}(2)$.

We have seen that description of differently $P$-oriented
geometries in terms  of spinor  fields $\eta$ and $\xi$ has  made hardly noticeable
distinction between these  two geometries much more
apparent  and intuitively appreciable
as connected with different types of spatial geometry indeed.

\section{Spinors $\xi $ and $\eta $ in cylindrical parabolic coordinates}

This coordinate system in initial $E_{3}$-space is defined by the relations
\begin{eqnarray}
x_{1} = {y^{2}_{1} - y^{2}_{2} \over 2 } \;\; ,\;\;
x_{2} =  y_{1} \; y_{2} \; \; ,  \;\; x_{3} = y_{3} \;\; ,\;\;
\;\;
y_{2} \in  [\; 0, + \infty\; ) \;\; , \;\; y_{1},\; y_{3} \in
 (\; - \infty , \; + \infty \;  ) \; .
\label{b2.1}
\end{eqnarray}

\noindent They can be  illustrated by the  figure

\vspace{5mm}

\unitlength=0.4mm
\begin{picture}(120,70)(-80,0)
\special{em:linewidth 0.4pt}
\linethickness{0.4pt}

\put(+10,+30){\vector(+1,0){100}}
\put(+112,+23){$y_{1}$}  \put(+60,0){\vector(0,+1){70}}
\put(+64,+67){$y_{2}$} \put(+20,+30){\line(0,+1){30}}
\put(+30,+30){\line(0,+1){30}} \put(+40,+30){\line(0,+1){30}}
\put(+50,+30){\line(0,+1){30}} \put(+70,+30){\line(0,+1){30}}
\put(+80,+30){\line(0,+1){30}} \put(+90,+30){\line(0,+1){30}}
\put(+100,+30){\line(0,+1){30}}
\end{picture}

\begin{center}
FIG. 10. Region $\;\;G(y_{1}, y_{2})^{ y_{3}} $
\end{center}


\noindent where domain $G(y_{1},y_{2})^{y_{3}}$
 (at arbitrary $y_{3}$)  ranging in the
half-plane $(y_{1},y_{2})$ covers the whole vector plane $(x_{1},x_{2})^{x_{3}}$.

The spinor $\xi $ of pseudo vector $\Pi_{3}$-model is given by

\begin{eqnarray}
\xi(y) =
\left ( \begin{array}{c}
\sqrt{(y^{2}_{3} + (y^{2}_{1} + y^{2}_{2})^{2} / 4)^{1/2} +
y_{3} }   e^{-i\gamma /2}  \\ [3mm]
\sqrt{(y^{2}_{3} + (y^{2}_{1} + y^{2}_{2})^{2} / 4)^1/2\; -
y_{3} }  e^{+i\gamma /2} \end{array} \right ),\;
e^{i\gamma /2} = {y_{1} + i y_{2} \over \sqrt{ y^{2}_{1} +
y^{2}_{2} }} \;   ,
\label{b2.2}
\end{eqnarray}

\noindent where the factor  $e^{i\gamma /2}$  runs through upper complex half-plane.
The one-to-one correspondence $\xi \longleftrightarrow
 (y_{1},y_{2},y_{3})$ is violated
at $x_{3}$-axis, at these peculiar point sets
$\Pi^{+}_{0}$  and $\Pi^{-}_{0}$  spinor looks as
\begin{eqnarray}
\xi ^{+}_{0} = \sqrt{+2 y_{3}} \left ( \begin{array}{c}
e^{-i\Gamma /2} \\ 0   \end{array} \right )  , \qquad
\xi ^{-}_{0} = \sqrt{-2 y_{3}}  \left ( \begin{array}{c}
0 \\ e^{+i\Gamma /2}   \end{array} \right ) \; ,
\label{b2.3a}
\end{eqnarray}

\noindent where a mute angle variable $\Gamma$  is used
$$
e^{+i\Gamma /2} = \lim_{ y_{1} \rightarrow  0 , y_{2} \rightarrow  0}
{ y_{1} + i y_{2} \over \sqrt{y^{2}_{1} + y^{2}_{2}}}\; .
$$

\noindent In the plane  $\Pi^{+\cap -}$ spinor  $\xi$ is  given by
\begin{eqnarray}
\xi ^{+\cap -}= {1 \over \sqrt{2}} \left ( \begin{array}{c}
 y_{1} - i y_{2} \\  y_{1} + i y_{2} \end{array}  \right ) \; .
\label{b2.3b}
\end{eqnarray}

For  a proper  vector  model,  formulas for $\eta $-spinor
look as (values $ +$ and $-$ taken by  symbol $\sigma$   there correspond
 to $x_{3}>0$  and $x_{3}<0$ half-spaces respectively)
\begin{eqnarray}
\eta^{\sigma } (y) =
\left (  \begin{array}{c}
\sqrt{ \sqrt{y^{2}_{3} + (y^{2}_{1} + y^{2}_{2})^{2}/ 4} -
{y^{2}_{1} + y^{2}_{2} \over 2}
}  \;  \sigma e^{-i\gamma /2}  \\[3mm]
\sqrt{  \sqrt{y^{2}_{3}  + (y^{2}_{1} + y^{2}_{2})^{2} /4} +
{y^{2}_{1} + y^{2}_{2} \over 2} }  \;\;  e^{-i\gamma /2}
\end{array} \right ) .
\label{b2.4a}
\end{eqnarray}

Now we are  to extend  the vector $E_{3}$ and $\Pi_{3}$ models to spinor ones.
To this end it is convenient  to employ  two new  variables $k$ and $\phi$ instead of
$y_{1}, y_{2}$:
$$
y_{1} = k \; \cos  \phi \;  , \;\; y_{2} = k \;\sin \phi \; ,
\;\; \phi \in [\; 0 , \; \pi\; ] \; ;
$$

\noindent in $x$-representation  we get to
$$
x_{1} = {k^{2} \over 2}\; cos 2\phi \; , \;\; x_{2} = {k^{2} \over 2 } \;
\sin 2\phi \; , \;\;\;  2\phi  \in  [ 0 , \; 2 \pi ]\;
$$

\noindent that leads to the following identification rule
in the set of boundary points of
the  domain  $G(y_{1}, y_{2})^{y_{3}}$
(covering vector spaces  $\Pi_{3}$  and $E_{3}$):

\vspace{5mm}
\unitlength=0.5mm
\begin{picture}(120,70)(-70,0)
\special{em:linewidth 0.4pt}
\linethickness{0.4pt}

\put(+10,+30){\vector(+1,0){100}}
\put(+110,+25){$y_{1}$}  \put(+60,0){\vector(0,+1){70}}
\put(+62,+70){$y_{2}$} \put(+20,+30){\line(0,+1){30}}
\put(+30,+30){\line(0,+1){30}} \put(+40,+30){\line(0,+1){30}}
\put(+50,+30){\line(0,+1){30}} \put(+70,+30){\line(0,+1){30}}
\put(+80,+30){\line(0,+1){30}} \put(+90,+30){\line(0,+1){30}}
\put(+100,+30){\line(0,+1){30}} \put(+60,+30){\circle{2}}
\put(+20,+30){\circle*{2}} \put(+30,+30){\circle*{2}}
\put(+40,+30){\circle*{2}}  \put(+50,+30) {\circle*{2}}
\put(+70,+30){\circle*{2}}  \put(+80,+30) {\circle*{2}}
\put(+90,+30){\circle*{2}}  \put(+100,+30){\circle*{2}}

\put(+60,+30){\oval(20,10)[b]}   \put(+60,+30){\oval(40,20)[b]}
\put(+60,+30){\oval(60,30)[b]}   \put(+60,+30){\oval(80,40)[b]}
\end{picture}

\begin{center}
FIG. 11. Region   $\;\;\;\; G(y_{1},y_{2})^{y_{3}}$
\end{center}

\vspace{5mm}

\noindent here identified points on the  boundary are connected by lines.

Bearing in mind that spinors $\xi(y)$ and $\eta (y)$
take on different, opposite in sign,
 values one can put forward the following simple way
 to construct extended (spinor) models
$\tilde{E}_{3}$ and $\tilde{\Pi}_{3}$:
it is sufficient  to double the range of $y_{2}$-variable:
$$
y_{2} \in  [\; 0 , \; + \infty  ) \;\;  \Longrightarrow  \;\;
   y_{2} \in  ( - \infty  , \;  + \infty  ) \; .
$$

\noindent  After  so  doing  the above factor
$e^{+i\gamma /2}$ will run through the full unit circle:

\vspace{7mm}
\unitlength=0.52mm
\begin{picture}(150,70)(-70,0)
\special{em:linewidth 0.4pt}
\linethickness{0.4pt}

\put(0,+67){$\tilde{G}(y_{1},y_{2})$}
\put(0,+30){\vector(+1,0){60}} \put(+58,+23){$y_{1}$}
\put(+30,0){\vector(0,+1){60}} \put(+34,+58){$y_{2}$}
\put(+30,+30){\oval(20,20)}  \put(+40,+30){\circle*{2}}
\put(+20,+35){\vector(0,-1){5}}  \put(+20,+25){\vector(0,+1){5}}


\put(+90,+67){$ e^{i\gamma /2} $}
\put(+90,+30){\vector(+1,0){60}}   \put(+120,0){\vector(0,+1){60}}
\put(+120,+30){\oval(20,20)}  \put(+130,+30){\circle*{2}}
\put(+110,+35){\vector(0,-1){5}}  \put(+110,+25){\vector(0,+1){5}}
\put(+80,+35){$\gamma =+2\pi$} \put(+80,+23){$\gamma =-2\pi$}
\end{picture}

\begin{center}
FIG. 12. $4\pi$ - continuity
\end{center}

\vspace{3mm}

\noindent
It is  important to note  the
substantial changing  in the identification rules at the
boundary set of  $G(y_{1},y_{2},y_{3})$ ---  now for extended domain
$\tilde{G}(y_{1},y_{2},y_{3})$  one  needs no special  rules at all.
Thus, in a sense,
the  domain $\tilde{G}(y_{1},y_{2},y_{3})$ appears to be simpler than
$G(y_{1},y_{2},y_{3})$.

Else one point  must be emphasized. The same  extended  set
$\tilde{G}(y_{1},y_{2},y_{3})$ is valid
to both spinor models $\xi(y)$ and $\eta (y)$.
This means that only chioce of the set with doubling dimension
 and identification rules
does not  determine in full  the  whole geometry of spinor spaces.
Specification of  their $P$-orientation  requires seemingly  additional  information
about this set. Unfortunately, this  point has not been clarified sufficiently.
Searching  the model under consideration for  some arguments
to state  those  distinctions
in rational way is the  main objective of the  present work.

Evidently, $P$-orientation  manifests itself in  explicitly different
spinor functions $\xi(y)$ and $\eta (y)$.
 Some qualitative  distinction between  these  spinor functions is revealed if one
follows  orientation of spinor $(\xi_{1},\xi_{2})$
and $(\eta_{1}, \eta_{2})$ while going from
$x^{+}_{3}$ -- half-space  to $x^{-}_{3}$ -- half-space.
Here the explaining diagrams  may be
given:

\vspace{10mm}

\unitlength=0,6mm
\begin{picture}(120,50)(-75,-30)
\special{em:linewidth 0.4pt}
\linethickness{0.4pt}

\put(-20,0){\vector(+1,0){40}} \put(+22,-5){$y_{1}$}
\put(0,-20){\vector(0,+1){40}} \put(-8,+20){$y_{2}$}

\put(0,0){\circle{16}}  \put(+8,0){\circle*{2}} \put(+8,0){\vector(0,+1){6}}
\put(-15,+5){$1$} \put(-15,-8){$2$}

\put(+45,+20){$\xi^{1}_{+,-}$}
\put(35,0){\line(+1,0){30}}  \put(50,-15){\line(0,+1){30} }
\put(50,0){\circle{16}}  \put(+58,0){\circle*{2}} \put(+58,0){\vector(0,-1){6}}
\put(+38,+5){$2$} \put(+38,-8){$1$}

\put(+85,+20){$\xi^{2}_{+,-}$}
\put(75,0){\line(+1,0){30}}  \put(90,-15){\line(0,+1){30} }
\put(90,0){\circle{16}}  \put(+98,0){\circle*{2}} \put(+98,0){\vector(0,+1){6}}
\put(+78,+5){$1$} \put(+78,-8){$2$}

\end{picture}

\begin{center}
FIG. 13.  ($\;\xi$ - model)
\end{center}

\vspace{+5mm}

\unitlength=0,6mm
\begin{picture}(120,50)(-75,-30)
\special{em:linewidth 0.4pt}
\linethickness{0.4pt}

\put(-20,0){\vector(+1,0){40}} \put(+22,-5){$y_{1}$}
\put(0,-20){\vector(0,+1){40}} \put(-8,+20){$y_{2}$}

\put(0,0){\circle{16}}  \put(+8,0){\circle*{2}} \put(+8,0){\vector(0,+1){6}}
\put(-15,+5){$1$} \put(-15,-8){$2$}

\put(+45,+20){$\eta^{1}_{+}$}
\put(35,0){\line(+1,0){30}}  \put(50,-15){\line(0,+1){30} }
\put(50,0){\circle{16}}  \put(+58,0){\circle*{2}} \put(+58,0){\vector(0,-1){6}}
\put(+38,+5){$2$} \put(+38,-8){$1$}

\put(+85,+20){$\eta^{2}_{+,-}$}
\put(75,0){\line(+1,0){30}}  \put(90,-15){\line(0,+1){30} }
\put(90,0){\circle{16}}  \put(+98,0){\circle*{2}} \put(+98,0){\vector(0,+1){6}}
\put(+78,+5){$1$} \put(+78,-8){$2$}

\put(+45,-23){$\eta^{1}_{-}$}
\put(35,-40){\line(+1,0){30}}  \put(50,-55){\line(0,+1){30} }
\put(50,-40){\circle{16}}  \put(+42,-40){\circle*{2}} \put(+42,-40){\vector(0,+1){6}}
\put(+38,-35){$1$} \put(+38,-48){$2$}

\end{picture}

\vspace{20mm}

\begin{center}
FIG. 14.   ($\;\eta$ - model)
\end{center}

\vspace{5mm}

Numbers 1 and 2 there correspond to  first (initial)
and second (additional) sub-space of the  whole
space with spinor structure.

Else one method to describe  spatial spinor $\xi (y)$ and
$\eta (y)$ with the  help of
coordinates $y_{i}$ is the  2-gradient:
\begin{eqnarray}
{\partial  \over \partial y_{1}}  \xi ^{1} =
{\xi ^{1} \over 2}  (
{\rho  \over a (a  + a_{3}) }\; y_{1} +
{ i \over \rho } \; y_{2}  ) \; , \qquad
{\partial  \over \partial y_{2}} \xi ^{1} =
{\xi ^{1} \over 2}  (
{\rho  \over a (a  + a_{3}) } \; y_{2}  -
{ i \over \rho } \; y_{1}   ) \; ,
\nonumber
\\
{\partial \over \partial y_{1}} \xi ^{2} =
{\xi ^{2}\over 2}  (
{\rho  \over a (a - a_{3}) } \; y_{1} -
{ i \over \rho } y_{2}  ) \;,\qquad
{\partial \over \partial y_{2}} \xi ^{2} =
{\xi ^{2} \over 2}  (
{\rho  \over a (a - a_{3}) }  y_{2}  +
{ i \over \rho } \; y_{1}  ) ,
\label{b2.5}
\end{eqnarray}
\begin{eqnarray}
{\partial \over \partial y_{1}} \eta ^{1} =
{\eta ^{1} \over  2} (
 - { y_{1} \over b}  +i
  {y_{2} \over \rho }  )   , \qquad
 {\partial \over \partial y_{2}}  \eta ^{1} =
{\eta ^{1} \over 2}  (
- { y_{2} \over b} - i  {y_{1} \over \rho }  )  , \qquad
\nonumber
\\
{\partial \over \partial y_{1}}  \eta ^{2} =
{\eta ^{2} \over 2}   (
+ { y_{1} \over b}  - i {y_{2} \over \rho }  ) , \qquad
{\partial \over \partial y_{2}} \eta ^{2} =
{\eta ^{2} \over 2} (
+ { y_{2} \over b}  +  i   {y_{1} \over \rho }  ) . \qquad
\label{b2.6}
\end{eqnarray}

\noindent
Formulas  (\ref{b2.5})   and (\ref{b2.6}) have no peculiarities
over complex plane $y_{1}+ i y_{2}$,
excluding  the origin point $0+i0$.
From (\ref{b2.5}),(\ref{b2.6}) it follows the
explicit form of derivatives with respect to
direction in $(y_{1},y_{2})$-plane:
\begin{eqnarray}
\nabla _{\vec{\nu }} \; \xi ^{1} = {\xi ^{1}\over 2}
\left [ {\rho \over a (a + a_{3}) }  (\vec{\nu }\; \vec{y}) +
{i \over \rho }  (\vec{\nu } \times \vec{y})\;  \right ]  ,
\nonumber
\\[2mm]
\nabla_{\vec{\nu }} \; \xi ^{2} = {\xi ^{2}\over 2} \left [
{\rho \over a (a - a_{3}) }   (\vec{\nu }\; \vec{y}) -
 {i \over \rho }   (\vec{\nu } \times \vec{y}) \right ]  ,
\label{b2.7}
\end{eqnarray}

\noindent and
\begin{eqnarray}
\nabla_{\vec{\nu }}  \; \eta ^{1} = {\eta ^{1}\over 2}
\left [-  {\vec{\nu }  \vec{y} \over b} +
{i \over \rho }  (\vec{\nu } \times  \vec{y})  \right ]  ,\qquad
\nabla_{\vec{\nu }} \;  \eta ^{2} = {\eta ^{2}\over 2}
\left [  {\vec{\nu }\; \vec{y})\over b} -
{i\over \rho } (\vec{\nu } \times \vec{y})  \right ]  ,
\label{b2.8}
\end{eqnarray}

\noindent where the notation is used:
\begin{eqnarray}
\vec{y} = ( y_{1}, y_{2} )  , \;\;
\vec{\nu } = ( \nu _{1}, \nu _{2} )  ,\;\;
(\vec{\nu } \; \vec{y}) =  \nu _{1} y_{1}+ \nu _{2} y_{2}
\; ,\;\;
 (\vec{\nu } \times  \vec{y}) =
 \nu _{1}  y_{2} - \nu _{2}  y_{1}  \; .
\nonumber
\end{eqnarray}

Relations  (\ref{b2.7}) and (\ref{b2.8}) can be considered
 alternatively  as basic equations  that
prescribe the explicit form of spinors $\xi (y)$ and $\eta (y)$ -- from which we had
started  in the beginning. Such understanding  of equations
of the type (\ref{b2.7}) and (\ref{b2.8})
 appears
to be interesting and  possibly fruitful. As for now, they do not look simple or
fundamental anyhow, however having been in
their infancy they do have  exiting mathematical
potential.

\vspace{7mm}

Let us examine  certain interesting properties
of the mapping $(x_{1},x_{2},x_{3}) \Longrightarrow \;
\tilde{G}(y_{1},y_{2},y_{3})$,
which look the same for both models $\tilde{E}_{3}$ and
$\tilde{\Pi}_{3}$.  For neighborhood of any point $ \vec{y}_{0}$

\vspace{10mm}
\unitlength=0.6mm
\begin{picture}(100,60)(-75,0)
\special{em:linewidth 0.4pt}
\linethickness{0.4pt}

\put(-10,+60){$\vec{y} = ( \vec{y}_{0} + \epsilon \;\vec{\nu})$}
\put(-10,+50){$\vec{\nu } = ( \cos  \phi , \;\sin \phi )$}

\put(+10,+30){\vector(+1,0){80}} \put(+90,+25){$y_{1}$}
\put(+50,20){\vector(0,+1){40}} \put(+52,+60){$y_{2}$}
\put(+50,+30){\circle*{2}} \put(+50,+30){\vector(+3,+2){15}}
\put(+45,+35){$\vec{y}_{0}$} \put(+65,+41){$\vec{\nu}$}
\put(+75,+45){\circle*{2}} \put(+75,+45){\vector(+3,+2){15}}
\put(+75,+45){\line(+1,0){15}} \put(+91,+53){$\vec{\nu}$}
\put(+70,+50){$\vec{y}_{0}$}
\end{picture}

\vspace{-13mm}

\begin{center}
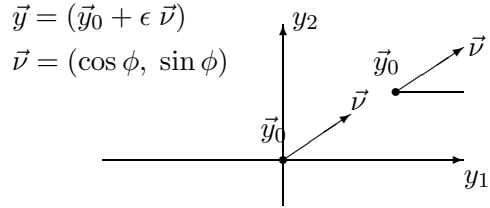
FIG. 15. Neighborhood of the point $ \vec{y}_{0}$
\end{center}

\vspace{+4mm}

\noindent in $x$-representation one gets
\begin{eqnarray}
x_{1} =  x^{0}_{1} + \epsilon  ( y^{0}_{1} \cos  \phi
- y^{0}_{2} \sin  \phi ) +  (\epsilon ^{2}/2)  \cos
2 \phi  \; ,
\nonumber
\\
x_{2} =  x^{0}_{2} + \epsilon (y^{0}_{1} \sin  \phi +
y^{0}_{2} \cos  \phi  ) + (\epsilon ^{2}/2) \sin  2\phi \; .
\nonumber
\end{eqnarray}

\noindent If $y^{0}_{1} = 0$ and $y^{0}_{2} = 0$, the first order terms   vanish
and
we  have
\begin{eqnarray}
x_{1}  = + (  \epsilon ^{2}/2)
\cos 2\phi   ,  \;\;\;   x_{2}  = + (\epsilon ^{2}/2) \sin 2\phi \;.
\nonumber
\end{eqnarray}

\noindent The latter means that in the
vicinity of $(0,0)$-point just the  angle $2\phi$
(in contrast to $\phi$-variable)  has  a first-hand geometrical sense.
In accordance with $\phi \in [0,2\pi ]$
(here  an extended $y_{1},y_{2})$-range  has been
presupposed) the variable $2\phi$ runs through  the  double interval  $[0, 4\pi ]$.
The part (sub-interval) $\phi \in [0, 2 \pi ]$
there corresponds to the first sheet and the
part $\phi \in [2\pi, 4 \pi ]$ -- to the second  one
of the 2-sheeted $(x_{1},x_{2})$-plane.
In all remaining points  the plane $y_{1} + y_{2} $,
first-hand geometrical  meaning of
$\phi$-variable follows from the formulas
$$
x_{1} = a^{0}_{1}  +  \epsilon  \;
\sqrt{(y^{0}_{1})^{2} + ( y^{0}_{2})^{2}} \; \cos
(\phi  + \Delta (y))  ,
$$
$$
x_{2} = a^{0}_{2} + \epsilon  \sqrt{(y^{0}_{1})^{2} +
(y^{0}_{2})^{2}} \; \sin (\phi  + \Delta (y))   ,
$$

\noindent where  $\Delta (y)$ is  defined by
\begin{eqnarray}
\cos \Delta (y) = { y^{0}_{1} \over \sqrt{(y^{0})^{2}  +
(y^{0})^{2}}}  \; , \qquad
\sin  \Delta (y) = {y^{0}_{2}\over \sqrt{ (y^{0})^{2}  +
(y^{0})^{2}}} \;  .
\nonumber
\end{eqnarray}

\noindent This means that
at all such points the variable $\phi$ ranging in
$[0,2\pi ]$-interval has ordinary geometrical sense.

The  property  just described can be reformulated as follows:
all points of $\tilde{E}_{3}$
and $\tilde{\Pi}_{3}$, different from $(0,0,x_{3})$,
are characterized  by  $2\pi$-neighborhoods
of directions, whereas  in the vicinity of all point $(0,0,x_{3})$
there exist $4\pi$--neighborhoods
of directions. Evidently,  that geometrical construction is well known
in the  complex variable function
theory as it concerns  2-sheeted complex plane.

Else one point may be  noticed. In all $2\pi$-points of the
extended space spinors $\xi(y)$ and
$\eta (y)$ are single-valued  functions of  spatial points
$(y_{1},y_{2},y_{3})$; whereas
in all $4\pi$-points (the whole axis $(0,0,x_{3})$) spinors
 are not single-valued functions --
they have discontinuity described by
the exponential factor $e^{\pm i \gamma/2 }$. As the  variable
$\gamma$ ranges  from $0$ to $4\pi$, we will have in all $4\pi$-points
\begin{eqnarray}
\xi (\gamma  = 0 )  = \xi (\gamma  = 4\pi ) \; , \;\;\;
\eta
(\gamma = 0 ) = \eta (\gamma  = 4\pi ) \; .
\nonumber
\end{eqnarray}

In other words,  spinor $\xi(y)$ and  are  $\eta (y)$
continuous in every point of the whole
space with respect to its direction set.
The latter may be characterized  symbolically as follows:
$$
2\pi \; \otimes \pi \;\;  \mbox{for} \;\;  2\pi-\mbox{points} \; ; \qquad
 4\pi \; \otimes \pi \;\;  \mbox{for} \;\; 4\pi-\mbox{points} \; .
 $$

In the following,  for the sets of discontinuity points we will  employ designation
$
R^{exp}, \; R^{\pm 1} $ and  $\tilde{R}^{exp}$,
where symbol of tilde  refers to extended  models. The domains
$R^{exp}, \;R^{\pm 1} $ are presented in initial vector models, in spinor models
there only domain $\tilde{R}^{exp}$ arises.
The latter can be illustrated by the diagram:

\vspace{7mm}
\unitlength=0.5mm
\begin{picture}(120,70)(-70,0)
\special{em:linewidth 0.4pt}
\linethickness{0.4pt}

\put(+10,+30){\vector(+1,0){100}}
\put(+110,+25){$y_{1}$}  \put(+60,0){\vector(0,+1){70}}
\put(+62,+70){$y_{2}$} \put(+20,+30){\line(0,+1){30}}
\put(+30,+30){\line(0,+1){30}} \put(+40,+30){\line(0,+1){30}}
\put(+50,+30){\line(0,+1){30}} \put(+70,+30){\line(0,+1){30}}
\put(+80,+30){\line(0,+1){30}} \put(+90,+30){\line(0,+1){30}}
\put(+100,+30){\line(0,+1){30}} \put(+60,+30){\circle{2}}
\put(+20,+30){\circle*{2}} \put(+30,+30){\circle*{2}}
\put(+40,+30){\circle*{2}}  \put(+50,+30) {\circle*{2}}
\put(+70,+30){\circle*{2}}  \put(+80,+30) {\circle*{2}}
\put(+90,+30){\circle*{2}}  \put(+100,+30){\circle*{2}}

\put(+60,+22){$ R^{exp.}$}
\put(+30,+22){$R^{\pm 1}$}
\put(+90,+22){$R^{\pm 1}$}
\end{picture}

\vspace{+2mm}

\begin{center}
FIG.16. $\;\; G(y_{1},y_{2})^{y_{3}} \;\; \mbox{for}
\;\; \Pi_{3}\;\; \mbox{and} \;\; E_{3}$
\end{center}
\unitlength=0.5mm
\begin{picture}(120,70)(-80,0)
\special{em:linewidth 0.4pt}
\linethickness{0.4pt}

\put(+10,+30){\vector(+1,0){100}}
\put(+110,+25){$y_{1}$}  \put(+60,0){\vector(0,+1){70}}
\put(+62,+70){$y_{2}$}
\put(+20,+30){\line(0,+1){30}}
\put(+30,+30){\line(0,+1){30}} \put(+40,+30){\line(0,+1){30}}
\put(+50,+30){\line(0,+1){30}} \put(+70,+30){\line(0,+1){30}}
\put(+80,+30){\line(0,+1){30}} \put(+90,+30){\line(0,+1){30}}
\put(+100,+30){\line(0,+1){30}} \put(+60,+30){\circle{2}}

\put(+20,+30){\line(0,-1){30}}
\put(+30,+30){\line(0,-1){30}} \put(+40,+30){\line(0,-1){30}}
\put(+50,+30){\line(0,-1){30}} \put(+70,+30){\line(0,-1){30}}
\put(+80,+30){\line(0,-1){30}} \put(+90,+30){\line(0,-1){30}}
\put(+100,+30){\line(0,-1){30}}

\put(+60,+22){$ R^{exp.}$}
\end{picture}

\begin{center}
FIG. 17.
 $\;\; \tilde{G}(y_{1},y_{2})^{y_{3}} \;\; \mbox{for}
\;\; \tilde{\Pi}_{3}\;\; \mbox{and} \;\; \tilde{…}_{3}$
\end{center}

Manifestation  of sets
$R^{exp}$ and $\tilde{R}^{exp}$ differ from each
other when $\gamma \in [0,2\pi ]$ and
$\gamma \in [0,4\pi ]$ respectively.
In view of such continuity properties of spinors
$\xi(y)$ and $\eta(y)$ one  may generalize
the concept of a point of spinor space

\unitlength=0.6mm
\begin{picture}(30,30)(-130,0)
\special{em:linewidth 0.4pt}
\linethickness{0.4pt}

\put(-90,+10){$\xi (\vec{x}) \rightarrow \xi (\vec{a}, \vec{m})
\; $}
\put(-90,-10){$\eta (\vec{x}) \rightarrow \eta (\vec{b},
\vec{m}) \;   $}

\put(0,0){\oval(20,20)} \put(0,0){\vector(+1,0){20}}
\put(0,0){\vector(-1,0){20}} \put(0,0){\vector(0,+1){20}}
\put(0,0){\vector(0,-1){20}} \put(0,0){\vector(+1,+1){10}}
\put(0,0){\vector(+1,-1){10}} \put(0,0){\vector(-1,+1){10}}
\put(0,0){\vector(-1,-1){10}}
\end{picture}

\vspace{10mm}

\begin{center}
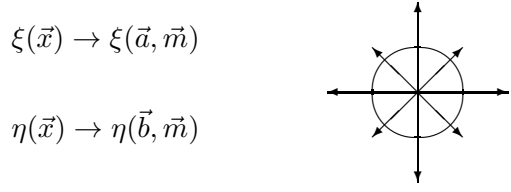
FIG. 18. Point of spinor space
\end{center}
\vspace{-10mm}

\vspace{15mm}

\noindent that is "a point of spinor space" \hspace{2mm}
is an  aggregate formed by the point
$\vec{x} - (y_{1},y_{2}, y_{3})$ as such and
by the direction set $\{ \vec{m}\}$ near the point.

And a final remark. In the spinor  space  models one  can readily
determine a  metric structure with the
help of ordinary metric tensor in  cylindrical parabolic coordinates
\begin{eqnarray}
dl^{2} = [\; dy^{2}_{3} \; + \; ( y^{2}_{1}\; + \; y^{2}_{2})
(dy^{2}_{1} \; + \; dy^{2}_{2} ) \; ] \;
\label{b2.11}
\end{eqnarray}

\noindent where coordinates  range in the extended domain,
covering initial vector space twice:
$$
\tilde{G}((y_{1},y_{2},y_{3}), \qquad
\; y_{i}  \;  \in  \; (- \infty , + \infty ) \; , \;\; i = 1,2,3 \; .
$$

The case of cylindrical parabolic coordinates  provides
us with important  tool to describe
spinor  spaces. In a sense, the structure of spaces
with spinor properties in  terms of these
coordinates looks simpler than of vector space -- compare
identification rules for  boundary
points. However it must be  mentioned else one time:
to distinguish between spinor models of
different $P$-type, the given   specification of $\tilde{G}(y_{1},y_{2},y_{3})$
 (geometrical dimension and  boundary identification)
  is not sufficient, and some additional
 mathematical  technique should  be  elaborated.

\section{Spinors  $\xi$  and  $\eta$  in parabolic coordinates}

In this Section we  are going to  examine  in spinor approach the  well-known parabolic
 coordinates.  They are defined  by the formulas
\begin{eqnarray}
x_{1} = y_{1}  y_{2} \; \cos  y_{3}\; , \;
   \;
x_{2} = y_{1}  y_{2}\;  \sin y_{3} \; , \;
x_{3} = {y^{2}_{1} - y^{2}_{2} \over 2}\;  , \;\;
 y_{1},y_{2} \in  [  0  , + \infty )\;  , \;\;
y_{3} \in  [  0 , 2 \pi   ]\qquad
\nonumber
\end{eqnarray}

\noindent with the  diagram
\vspace{+3mm}

\unitlength=0.6mm
\begin{picture}(50,50)(-80,0)
\special{em:linewidth 0.4pt}
\linethickness{0.4pt}

\put(0,0){\vector(+1,0){40}} \put(+40,-5){$y_{1}$}
\put(0,0){\vector(0,+1){40}} \put(+2,+40){$y_{2}$}
\put(0,0){\vector(+1,+1){40}}
\put(+42,+32){$E^{+}_{3} \cap E^{-}_{3}$}
\put(-18,+20){$E^{+}_{0} \rightarrow $}
\put(+20,+4){$E^{-}_{0} \downarrow $}
\end{picture}
\vspace{5mm}

\begin{center}
FIG. 19. Parabolic coordinates
\end{center}

Spatial spinor $\eta$ of the properly vector model is  given by
\begin{eqnarray}
\eta ^{+}(y) =  {1 \over \sqrt{2}}  \left ( \begin{array}{c}
 (y_{1} - y_{2})\;\;  e^{-iy_{3}/2} \\
 ( y_{1} + y_{2})  \;\; e^{+iy_{3}/2}
\end{array} \right )  , \qquad
\nonumber
\\
\eta ^{-}(y) = {1 \over \sqrt{2}} \left ( \begin{array}{lr}
 (y_{2} - y_{1}) &(- e^{-iy_{3}/2} ) \\
 (y_{2} + y_{1}) & e^{+iy_{3}/2}
\end{array} \right ) \; . \;
\label{b3.2a}
\end{eqnarray}

\noindent Spinors  $\eta^{\pm}_{0}, \eta ^{+\cap -} $ look as follows
\begin{eqnarray}
\eta ^{+}_{0}= { y_{1} \over \sqrt{2}} \left ( \begin{array}{c}
e^{-i\Gamma /2} \\ e^{+i\Gamma /2} \end{array} \right )  ,\;
\eta ^{-}_{0} = {y_{1} \over \sqrt{2}} \left ( \begin{array}{c}
- e^{-i\Gamma /2} \\  e^{+i\Gamma /2} \end{array} \right )  ,\;
\eta ^{+\cap -} = \left ( \begin{array}{c}
0 \\ \sqrt{2} y e^{+iy_{3}/2} \end{array} \right )  .
\nonumber
\end{eqnarray}

\noindent where  $\Gamma $ is a mute  variable,
the notation $y_{1}=y_{2}=y$ is used for
the  plane $x_{3}=0$.

As for  pseudo vector  model $\Pi_{3}$ we will have
\begin{eqnarray}
\xi (y) =   \left ( \begin{array}{c}
y_{1} \; e^{-iy_{3}/2} \\ y_{2} \; e^{+iy^{3}/2}  \end{array} \right )\; .
\label{b3.3a}
\end{eqnarray}

On comparing  (\ref{b3.3a}) with  definition of spatial spinor
$$
\xi (y) =   \left ( \begin{array}{c}
N \; e^{-i\gamma /2} \\ M  \; e^{+i\gamma /2}  \end{array} \right )\; .
$$

\noindent  we  immediately arrive at
\begin{eqnarray}
y_{1} = N \; ,\; y_{2} = M \; , \; y_{3} = \gamma \; .
\label{b3.3b}
\end{eqnarray}

\noindent In other words,  parabolic coordinates  $( y_{1},y_{2},y_{3})$
just coincide with  $(N, M, \gamma)$ introduced
in Sec. 1 at defining the concept of spinor $\eta$.

\vspace{3mm}
Now let us  outline some details of continuity property of spinors
$\xi$ and $\eta$.

\vspace{+32mm}
\unitlength=0.6mm
\begin{picture}(100,60)(-70,0)
\special{em:linewidth 0.4pt}
\linethickness{0.4pt}
\vspace{-5mm}

\put(+10,+50){\vector(+1,0){80}}
\put(+90,+45){$x_{2}$}   \put(+50,0){\vector(0,+1){100}}
\put(+42,+100){$x_{3}$}  \put(+80,65){\vector(-2,-1){60}}
\put(+13,+35){$x_{1}$}
\put(+5,+70){$R^{\pm}\;\; \rightarrow \;\;$}
\put(+5,+30){$R^{\pm}\;\; \rightarrow \;\;$}

\put(+48,+90){\line(-2,-1){30}}   \put(+48,+80){\line(-2,-1){30}}
\put(+48,+70){\line(-2,-1){30}}   \put(+48,+60){\line(-2,-1){30}}
\put(+48,+40){\line(-2,-1){30}}   \put(+48,+30){\line(-2,-1){30}}
\put(+48,+20){\line(-2,-1){30}}   \put(+48,+10){\line(-2,-1){30}}
\put(+51,+70){$ \leftarrow \;\; R^{exp.}$}
\put(+51,+5){$\leftarrow R^{exp.}$}
\end{picture}
\vspace{3mm}

\begin{center}
FIG. 20. $R^{\pm1}, R^{exp}$ in $x$-representation

\end{center}

and

\vspace{5mm}
\unitlength=0.6mm
\begin{picture}(110,100)(-60,0)
\special{em:linewidth 0.4pt}
\linethickness{0.4pt}

\put(+60,+50){\vector(+1,0){70}}   \put(+130,+45){$y_{2}$}
\put(+60,+50){\vector(0,+1){50}}   \put(+53,+100){$y_{3}$}
\put(+60,+50){\vector(-1,-1){40}}  \put(+25,+10){$y_{1}$}
\put(+45,+85){$ + 2 \pi $}
\put(+55,+70){\circle*{2}}         \put(+55,+35){\circle*{2}}
\put(+55,+70){\line(0,-1){35}}     \put(+55,+72){$R^{\pm}$}
\put(+55,+30){$R^{\pm}$}
\put(+70,+80){\circle*{2}}         \put(+70,+45){\circle*{2}}
\put(+70,+80){\line(0,-1){35}}     \put(+72,+80) {$R^{\pm}$}

\put(+60,+85){\line(+1,0){60}}   \put(+60,+85){\line(-1,-1){30}}
\put(+30,+55){\line(+3,+1){90}}  \put(+120,+50){\line(0,+1){35}}
\put(+30,+20){\line(+3,+1){90}}  \put(+30,+20){\line(0,+1){35}}

\put(+72,+45){$R^{\pm}$}
\put(+10,+40){$R^{exp.}\;\Longrightarrow $}
\put(+110,+70){$\Longleftarrow \; R^{exp.}$}
\end{picture}

\begin{center}
FIG 21.  $R^{\pm1}, R^{exp}$ in $y$-representation

\end{center}

\vspace{5mm}

\noindent Transition to  extended space is achieved by doubling the above domain
$G(y) \Longrightarrow \tilde{G}(y)$

\vspace{10mm}

\unitlength=0.5mm
\begin{picture}(140,100)(-60,0)
\special{em:linewidth 0.4pt}
\linethickness{0.4pt}

\put(+60,+50){\vector(+1,0){70}}  \put(+130,+45){$y_{2}$}
\put(+60,0){\vector(0,+1){100}}    \put(+53,+100){$y_{3}$}

\put(+60,+50){\vector(-1,-1){40}}
\put(+25,+10){$y_{1}$}     \put(+45,+85){$ +2 \pi $}
\put(+45,+15){$-2 \pi $}   \put(+55,+70){\circle*{2}}
\put(+55,0){\circle*{2}}   \put(+55,+70){\line(0,-1){70}}
\put(+70,+80){\circle*{2}} \put(+70,+10){\circle*{2}}
\put(+70,+80){\line(0,-1){70}}
\put(+10,+40){$R^{exp.}\;\Longrightarrow $}

\put(+60,+85){\line(+1,0){60}}   \put(+60,+85){\line(-1,-1){30}}
\put(+30,+55){\line(+3,+1){90}}  \put(+30,+20){\line(+3,+1){90}}

\put(+120,+15){\line(0,+1){70}}  \put(+30,+20){\line(+3,+1){90}}
\put(+30,-15){\line(+3,+1){90}}  \put(+30,-15){\line(0,+1){70}}

\put(+90,+60){$\Longleftarrow \; R^{exp.}$}
\put(+10,0){$R^{exp.}\;   \Longrightarrow $}
\put(+90,+30){$\Longleftarrow \; R^{exp.}$}

\end{picture}

\vspace{11mm}

\begin{center}
FIG. 22.  Region   $\;\tilde{G}(y)$
\end{center}

\vspace{+5mm}

In parabolic  coordinates  spatial metrics is
\begin{eqnarray}
dl^{2} = [\; (y^{2}_{1} \;+ \; y^{2}_{2} ) \;
( dy^{2}_{1} + dy^{2}_{2})\; +\;
y^{2}_{1}\; y^{2}_{2}\; d y^{2}_{3}\; ] \;
\nonumber
\end{eqnarray}

\noindent  or
\begin{eqnarray}
d l^{2} =  (M^{2} + N^{2} )\; ( dM^{2} + dN^{2} ) + M^{2} N^{2}\;
d \gamma ^{2}  \; .
\nonumber
\end{eqnarray}

One final remark in this Section. You do not need  to employ  necessarily the
domain $\tilde{G}(y)$ described  above

\vspace{+15mm}
\unitlength=0.45mm

\begin{picture}(140,50)(-70,-20)
\special{em:linewidth 0.4pt}
\linethickness{0.4pt}

\put(0,0){\vector(+1,0){40}}   \put(+45,-5){$y_{1}$}
\put(0,0){\vector(0,+1){40}}  \put(-5,+40){$y_{2}$}
\put(0,0){\line(1,1){40}}
\put(0,+10){\line(1,1){30}}
\put(+10,0){\line(1,1){30}}
\put(0,+20){\line(1,1){20}}
\put(+20,0){\line(1,1){20}}

\put(+55,0){$\oplus$}

\put(70,0){\vector(+1,0){85}}     \put(130,0){\circle*{2}}  \put(90,0){\circle*{2}}
\put(155,-8){$y_{3}$}               \put(110,0){\circle*{2}}
\put(130,+5){$+2\pi$}  \put(90,5){$-2\pi$}

\end{picture}

\begin{center}
FIG. 23.   $\;\tilde{G}(y)$
\end{center}

\noindent
where the key role  in extending procedure  is assigned
to angle variable $y_{3}= \gamma$. Alternatively,
instead another (alternative, simple and symmetrical)  possibility exists

\vspace{5mm}

\vspace{+10mm}
\unitlength=0.38mm

\begin{picture}(140,50)(-100,-20)
\special{em:linewidth 0.4pt}
\linethickness{0.4pt}

\put(-40,0){\vector(+1,0){80}}   \put(+45,-5){$y_{1}$}
\put(0,-40){\vector(0,+1){80}}  \put(-5,+40){$y_{2}$}
\put(0,0){\line(1,1){40}}
\put(0,+10){\line(1,1){30}}
\put(+10,0){\line(1,1){30}}
\put(0,+20){\line(1,1){20}}
\put(+20,0){\line(1,1){20}}

\put(0,0){\line(-1,-1){40}}
\put(0,-10){\line(-1,-1){30}}
\put(-10,0){\line(-1,-1){30}}
\put(0,-20){\line(-1,-1){20}}
\put(-20,0){\line(-1,-1){20}}

\put(+55,0){$\oplus$}

\put(70,0){\vector(+1,0){85}}     \put(130,0){\circle*{2}}
\put(155,-8){$y_{3}$}               \put(110,0){\circle*{2}}
\put(130,+5){$+2\pi$}

\end{picture}

\vspace{7mm}
\begin{center}
FIG. 24.    $\;\tilde{G}(y)$
\end{center}

\vspace{5mm}

Thus, various  domains  $\tilde{G}(y)$ are acceptable for  correct
parameterizations  of spinor spaces, and you may choose  any for  reason of convention.

\section{Connection between  $\xi $  and  $\eta $ models}

Two models of spinors spaces  with respect
to $P$-orientation are grounded  on different
mappings  $\xi$   and  $\eta $  defined over  the same extended domain
$\tilde{G}(y_{i})$.  The natural question is:
how are  these two  maps  connected to each others. An
answer  can be  found on comparing the formulas  for $\xi $  and $\eta $:
\begin{eqnarray}
\eta (y) = {1 \over \sqrt{2}} \left ( \begin{array}{c}
(y_{1} - y_{2}) \; e^{-i y_{3}/2}  \\
(y_{1} + y_{2}) \; e^{+i y_{3}/2}
\end{array} \right ) \; , \qquad
\xi (y) = \left ( \begin{array}{c}
y_{1} \; e^{-i y_{3}/2}   \\   y_{2} \; e^{+i y_{3}/2}
\end{array} \right )\; . \qquad
\label{b4.1}
\end{eqnarray}

\noindent From (\ref{b4.1}) we   immediately arrive at
\begin{eqnarray}
\eta _{1} = { \xi _{1} - \xi _{2}^{*} \over \sqrt{2}} ,
\eta_{2}  = {\xi _{1}^{*} + \xi _{2} \over \sqrt{2}} \; ,\;\;
\eta = {1 \over \sqrt{2}} ( \xi - i \; \sigma ^{2} \xi ^{*} ) \; . \qquad
\label{b4.2}
\end{eqnarray}

\noindent Inverse to  (\ref{b4.2}) looks as
\begin{eqnarray}
\xi _{1} = { \eta _{1} + \eta _{2}^{*} \over \sqrt{2}} \; ,\;\;\;
\xi _{2} = { \eta _{2} - \eta _{1}^{*} \over \sqrt{2}} \; ,\;\;\;
\xi = {1 \over \sqrt{2}} \; ( \eta \;-\; i \;\sigma ^{2} \eta ^{*} )
\; . \qquad
\label{b4.3}
\end{eqnarray}

\noindent
In fact,  the formulas (\ref{b4.2}) and (\ref{b4.3})
are not coordinate-dependent -- one may obtain them  with the
use of any other coordinate system.
As for (\ref{b4.2}) and (\ref{b4.3})
there are two points that deserve special attention:

1) complex conjugation enters them explicitly which  correlates with the  change
in orientation properties of the  models;

2) spinors $\xi $   and  $i \sigma ^{2} \xi ^{*}$ (as well as  $\eta$
and $i \sigma ^{2} \eta ^{*}$)  provide us
with  non-equivalent representations of the extended
 unitary group $\tilde{SU}(2)$.

\section{Spatial spinors in  spherical  coordinates}

\hspace{7mm}
In this  Section we  will examine  in spinor approach
the most commonly encountered system of  spherical
coordinates. These are defined  by
\begin{eqnarray}
x_{1}= y_{1} \sin  y_{2}  \cos  y_{3}, \;
x_{2}= y_{1} \sin  y_{2}  \sin  y_{3}  ,
x_{3}= y_{1} \cos  y_{2}   \;,
\nonumber
\\
y_{1} \in  [ 0, + \infty  ) \; ,\; y_{2} \in  [ 0, + \pi  ]\; ,\;
y_{3} \in  [ 0, +2\pi  ]\; .
\label{b5.1}
\end{eqnarray}

\noindent Spinor $\eta(y)$ of pseudo vector  model $\Pi_{3}$
is  given by
\begin{eqnarray}
\xi  =   \left ( \begin{array}{c}
\sqrt{y_{1}(1 + \cos y_{2}) } \; e ^{-iy_{3}/2}  \\
\sqrt{y_{1}(1 - \cos y_{2})} \; e ^{+iy_{3}/2} \end{array} \right )\;,\qquad
\xi ^{+\cap -} = \sqrt{ y_{1} } \left ( \begin{array}{c}
e^{-iy_{3}/2 } \\ e^{+iy_{3}/2 } \end{array} \right ) \; ,
\label{b5.2}
\\
\xi ^{+}_{0}= \sqrt{2y_{2} } \left ( \begin{array}{c}
e^{-i\Gamma /2} \\ 0    \end{array} \right ) \; ,\qquad
\xi ^{-}_{0} = \sqrt{2y_{1} }   \left ( \begin{array}{c}
0 \\ e^{+i\Gamma /2}  \end{array} \right )\; , \;
\;\; ( \Gamma  = y_{3} )\; .
\end{eqnarray}

\noindent In turn, spinor $\eta(y)$ of properly vector model $E_{3}$
 is defined according to
\begin{eqnarray}
\eta  = \left ( \begin{array}{lr}
\sqrt{y_{1}(1 - \sin  y_{2})} & (\sigma  e^{-iy_{3}/2} \\
\sqrt{y_{1}(1 + \sin  y_{2})} &  e^{+iy_{3}/2}
\end{array} \right ) \; ,\qquad \qquad \qquad
\nonumber
\\
\eta ^{+\cap -} = \sqrt{ y_{1} } \left ( \begin{array}{c}
0         \\  \sqrt{2 y_{1} } e^{+iy_{3}/2}  \; ,
\end{array} \right ) \; , \;
\;
\eta ^{+}_{0} = \left ( \begin{array}{r}
e^{-i\Gamma /2} \\ e^{+i\Gamma /2}    \end{array} \right ) \; ; \;\;
\eta ^{-}_{0} = \left ( \begin{array}{r}
- e^{-i\Gamma /2} \\ e^{+i\Gamma /2}  \end{array} \right )\;   .
\label{b5.3}
\end{eqnarray}

\noindent
Discontinuity properties of  these spinors  may be characterized by the  diagram

\vspace{6mm}

\unitlength=0.6mm
\begin{picture}(100,50)(-90,+0)
\special{em:linewidth 0.4pt}
\linethickness{0.4pt}

\put(0,+10){\vector(+1,0){70}}
\put(+63,+5){$y_{3}$}      \put(+10,0){\vector(0,+1){50}}
\put(+3,+48){$y_{2}$}      \put(+10,+30){\line(+1,0){40}}
\put(+2,+30){$+\pi$}       \put(+50,+10){\line(0,+1){20}}
\put(+48,+5){$+2\pi$}      \put(+10,+30){\line(+1,0){40}}
\put(+30,+32){$R^{exp.}$}
\put(+53,+20){$\leftarrow R^{\pm 1}$} \put(+30,+3){$R^{exp.}$}
\put(-10,+20){$R^{\pm1} \rightarrow$ }
\end{picture}

\begin{center}
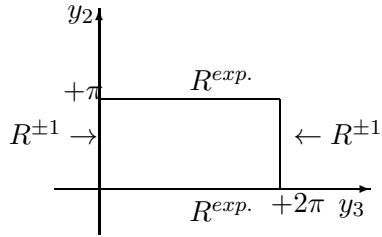
FIG.  25. Spinor discontinuity in spherical  coordinates
\end{center}

\vspace{5mm}

\noindent
Evidently,  transition to  extended models can be realized  through
formal doubling the range  of angle variable $y_{3}$ (
in the following we will use the more common
notation $y_{1} = r , y_{2} = \theta  ,  y_{3} = \phi  )$
\begin{eqnarray}
\tilde{G}(r,\theta ,\phi ) =  \{ \;
r\in  [ 0 , + \infty  ) \; ,\;
\theta  \in  [ 0 , + \pi ] ,
\phi  \in  [-2\pi , + 2\pi  ]  \;  \}  \; .
\label{b5.4}
\end{eqnarray}

Now let us  discuss  some  alternative  variants of extended  domain $\tilde{G}$
that might be   used for  covering spinor spaces.
By way of illustration, the  most natural and
symmetrical  possibility of this  type is to extend the range of radial variable:
\begin{eqnarray}
\tilde{G}'(r,\theta ,\phi ) = \{  \; r \in  (- \infty , + \infty  ) \; ,\;
\theta \in  [ 0 , + \pi ] , \phi  \in  [- \pi  ,- \pi ] \; \}  \; .
\label{b5.5}
\end{eqnarray}

\noindent
To prove it, let us  turn again to the above expression for $\xi$
\begin{eqnarray}
\xi (r,\theta ,\phi ) =
\left ( \begin{array}{ll}
\sqrt{1 + \cos  \theta }\; & (\; \sqrt{ r\; e^{i\phi } }\; )^{*} \\
\sqrt{1 - \cos \theta  } \;& \; (\; \sqrt{ r \;e^{i\phi } } \;)
\end{array} \right )  \; .
\label{b5.6}
\end{eqnarray}

\noindent This  function is  considered over   the old domain
$\tilde{G}(r,\phi , \theta)$:
\vspace{5mm}

\unitlength=0.4mm
\begin{picture}(100,100)(-90,+0)
\special{em:linewidth 0.4pt}
\linethickness{0.4pt}

\put(0,+50){\vector(+1,0){100}}  \put(+100,+45){$r$}
\put(+10,+0){\vector(0,+1){100}} \put(+5,+100){$\phi $}

\put(+10,+70){\line(+1,0){80}}  \put(-5,+70){$+\pi$}
 \put(+10,+70){\circle*{2}} \put(+95,+60){$A$}

\put(+10,+90){\line(+1,0){80}}  \put(-7,+90){$+2\pi$}
\put(+90,+80){$A'$} \put(+10,+90){\circle*{2}}

\put(+10,+30){\line(+1,0){80}} \put(-5,+30){$-\pi$}
\put(+90,+20){$B'$} \put(+10,+30){\circle*{2}} \put(+10,+10){\circle*{2}}

\put(+10,+10){\line(+1,0){80}} \put(-7,+10){$-2\pi$}
\put(+90,+40){$B$}

\put(+30,+10){\line(0,+1){80}}
\put(+50,+10){\line(0,+1){80}}
\put(+70,+10){\line(0,+1){80}}

\end{picture}

\begin{center}
FIG.  26. $(\phi,r)$ region of spinor space
\end{center}

\noindent where  vertical lines   have  joined  identified points of the
boundary set.
Taking into account  the equality
$
r \; e^{\pm (\phi \pm \pi )} = (- r )\; e^{\pm i\phi }\; ,
$
and  allowing for  connection between
$(A , A' )$   and $( B , B')$ sub-sets
$$
\phi  ^{A'} = ( \phi ^{A} + \pi  ) \; , \qquad
\;\phi ^{B'} = ( \phi ^{B} - \pi  ) \;,
$$

\noindent we readily  arrive  at two relations
\begin{eqnarray}
\xi  ( r , \theta  , \phi ^{A'})=
\xi ( - r , \theta  , \phi ^{A} ) \; , \qquad
 \xi  ( r ,\theta , \phi ^{B'}) = \xi (- r , \theta  , \phi ^{B} ) \; ,
\label{b5.7}
\end{eqnarray}

\noindent which  provide us with possibility
to employ the following $(r, \phi)$-domain:
\vspace{5mm}

\unitlength=0.4mm
\begin{picture}(100,60)(-70,+0)
\special{em:linewidth 0.4pt}
\linethickness{0.4pt}

\put(+0,+30){\vector(+1,0){100}} \put(+100,+25){$r$}
\put(+50,+0){\vector(0,+1){60}}  \put(+43,+55){$\phi $}
\put(+10,+50){\line(+1,0){80}}   \put(+90,+40){$A$}
\put(+10,+10){\line(+1,0){80}}   \put(+90,+20){$B$}
\put(+51,+51){$+\pi $}           \put(+38,+5){$-\pi $}
\end{picture}

\begin{center}
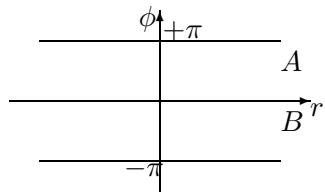
FIG.  27. Alternative $(\phi,r)$ region of spinor space

\end{center}

\vspace{5mm}
\vspace{5mm}

\noindent Observing identified points  on
 the initial diagram ($L \equiv L', F \equiv F'$
and so on )

\vspace{8mm}

\unitlength=0.5mm
\begin{picture}(100,100)(-90,+0)
\special{em:linewidth 0.4pt}
\linethickness{0.4pt}

\put(0,+50){\vector(+1,0){100}}  \put(+100,+45){$r$}
\put(+10,+0){\vector(0,+1){100}} \put(+1,+100){$\phi $}

\put(+10,+70){\line(+1,0){80}}  \put(-2,+70){$+\pi$}
\put(+10,+71){\line(+1,0){80}}
\put(+30,+72){$F$}    \put(+50,+72){$G$}
\put(+30,+63){$F'$}    \put(+50,+63){$G'$}

\put(+30,+32){$H$}    \put(+50,+32){$D$}
\put(+10,+31){\line(+1,0){100}}
\put(+30,+23){$H'$}    \put(+50,+23){$D'$}

\put(+30,+92){$L$}    \put(+50,+92){$M$}
\put(+30,+3){$L'$}    \put(+50,+3){$M'$}

 \put(+10,+70){\circle*{2}} \put(+90,+60){$A$}

\put(+10,+90){\line(+1,0){80}}  \put(-4,+90){$+2\pi$}
\put(+90,+80){$A'$} \put(+10,+90){\circle*{2}}

\put(+10,+30){\line(+1,0){80}} \put(-2,+30){$-\pi$}
\put(+90,+20){$B'$} \put(+10,+30){\circle*{2}}

\put(+10,+10){\line(+1,0){80}} \put(-4,+10){$-2\pi$}
\put(+90,+40){$B$}

\put(+30,+10){\line(0,+1){80}}
\put(+50,+10){\line(0,+1){80}}
\put(+70,+10){\line(0,+1){80}}

\end{picture}

\begin{center}
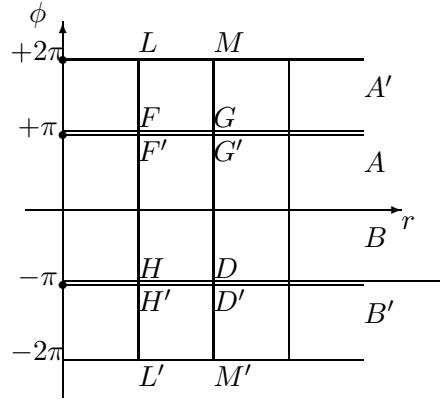
FIG.  28. Identification in $(\phi,r)$ region
\end{center}

\vspace{5mm}

\noindent
you can  easily  derive  identification rules  on the  new diagram for $\tilde{G}'$:

\vspace{5mm}

\unitlength=0.6mm
\begin{picture}(100,60)(-80,+0)
\special{em:linewidth 0.4pt}
\linethickness{0.4pt}

\put(+0,+30){\vector(+1,0){100}} \put(+100,+25){$r$}

\put(+0,+31){\line(+1,0){50}}  \put(+0,+29){\line(+1,0){50}}
\put(+10, +32){$G$} \put(+30, +32){$F$}
\put(+10, +22){$D'$} \put(+30, +22){$H'$}

\put(+50,+0){\vector(0,+1){60}}  \put(+45,+55){$\phi $}
\put(+70,+51){$F'$}    \put(+90,+51){$G'$}
\put(+70,+4){$H$}    \put(+90,+4){$D$}

\put(+10,+51){$M$}    \put(+30,+51){$L$}
\put(+10,+4){$M$}    \put(+30,+4){$L$}

\put(+0,+50){\line(+1,0){100}}   \put(+95,+40){$A$}
\put(+0,+10){\line(+1,0){100}}   \put(+95,+20){$B$}
\put(+50,+50){$+\pi $}           \put(+40,+5){$-\pi $}

\put(+10,+10){\line(0,+1){40}}   \put(+30,+10){\line(0,+1){40}}
\put(+70,+10){\line(0,+1){40}}   \put(+90,+10){\line(0,+1){40}}

\end{picture}

\begin{center}
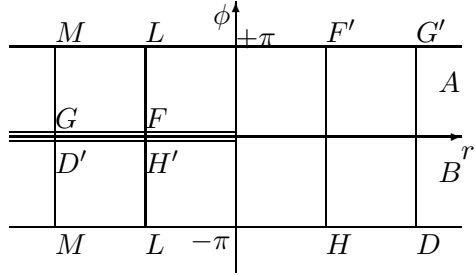
FIG.  29.  Identification in alternative $(\phi,r)$ region
\end{center}

\vspace{5mm}

\noindent
Transformation of the  domain  $\tilde{G}$  into   $\tilde{G}'$  can be illustrated by the
symbolic  relation
\begin{eqnarray}
[  R^{+}  \times  ( A'  + A + B + B' ) ]
\sim
 [ ( R^{+} +  R^{-} ) \times ( A + B ) ] .
\label{b5.8}
\end{eqnarray}

Needless to say that  two  domains $\tilde{G}$  and  $\tilde{G}'$   just  indicated
are not the  only  possible. For  example,  taking  into account  identities
\begin{eqnarray}
r \; e^{\pm i(\phi  \pm \pi )} = (- r ) \; e ^{\pm i\phi } \; ,
\qquad
 r\; e ^{\pm (\phi \pm 3\pi )}  = (- r ) \; e ^{\pm i\phi } \; ,
\nonumber
\end{eqnarray}

\noindent and  relationships between  $(A , B )$  and $(A', B')$
$$
\phi ^{B} = ( \phi ^{A} - \pi  )\; , \qquad  \phi ^{B'} =
( \phi ^{A'} - 3 \pi  ) \;
$$

\noindent we  readily produce the  formulas
\begin{eqnarray}
 \xi (r , \theta , \phi ^{B}) = \xi (- r, \theta , \phi ^{A}) \; , \qquad
  \xi  ( r , \theta  , \phi ^{B'}) = \xi (- r, \theta  , \phi ^{A'})
 \; .
\label{b5.9}
\end{eqnarray}

\noindent
They  mean that  instead of

\vspace{8mm}

\unitlength=0.49mm
\begin{picture}(100,100)(-90,+0)
\special{em:linewidth 0.4pt}
\linethickness{0.4pt}

\put(0,+50){\vector(+1,0){100}}  \put(+100,+45){$r$}
\put(+10,+0){\vector(0,+1){100}} \put(+5,+100){$\phi $}

\put(+10,+70){\line(+1,0){80}}  \put(-2,+70){$+\pi$}
\put(+10,+51){\line(+1,0){80}}
\put(+30,+52){$P$}    \put(+50,+52){$S$}
\put(+30,+43){$P$}    \put(+50,+43){$S$}

\put(+30,+32){$F$}    \put(+50,+32){$G$}
\put(+10,+31){\line(+1,0){80}}
\put(+30,+23){$F$}    \put(+50,+23){$G$}

\put(+30,+92){$L$}    \put(+50,+92){$M$}
\put(+30,+3){$L'$}    \put(+50,+3){$M'$}

 \put(+10,+70){\circle*{2}} \put(+90,+60){$A$}

\put(+10,+90){\line(+1,0){80}}  \put(-4,+90){$+2\pi$}
\put(+90,+80){$A'$} \put(+10,+90){\circle*{2}}

\put(+10,+30){\line(+1,0){80}} \put(-2,+30){$-\pi$}
\put(+90,+20){$B'$} \put(+10,+30){\circle*{2}}

\put(+10,+10){\line(+1,0){80}} \put(-4,+10){$-2\pi$}
\put(+90,+40){$B$}

\put(+30,+10){\line(0,+1){80}}
\put(+50,+10){\line(0,+1){80}}
\put(+70,+10){\line(0,+1){80}}

\end{picture}

\begin{center}
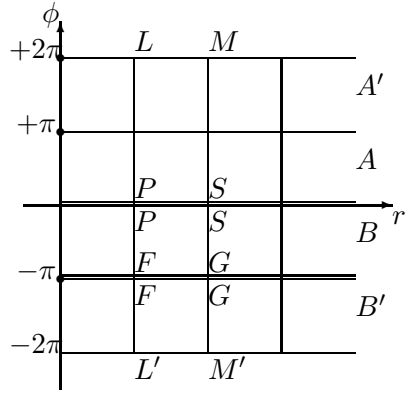
FIG.  30. Else one transformation of $(\phi,r)$  region
\end{center}

\vspace{1mm}
\noindent you may  use yet  another set   $\tilde{G}''$
$$
\bar{G}'' (r,\theta ,\phi ) =  \{
 r \in  (- \infty  ,+ \infty  ) ,\;
  \theta  \in  [ 0 , + \pi  ] ,
 \phi  \in  [ 0 , +2\pi  ]  \}  ;
$$
$$
[ R^{+}  \times  ( A'  + A + B + B' ) ]
\sim  [ ( R^{+} + R^{-} )  \times  ( A + A')   ]
$$

\noindent with identification rules as follows

\vspace{8mm}

\unitlength=0.55mm
\begin{picture}(100,60)(-80,+0)
\special{em:linewidth 0.4pt}
\linethickness{0.4pt}

\put(-10,+10){\vector(+1,0){120}} \put(+110,+5){$r$}

\put(+0,+31){\line(+1,0){50}}  \put(+0,+29){\line(+1,0){50}}
\put(+10,+2){$F$} \put(+30,+2){$G$}
\put(+10,+22){$S$} \put(+30,+22){$P$}
\put(+10,+32){$L'$} \put(+30,+32){$M'$}
\put(+10,+51){$F$} \put(+30,+51){$G$}
\put(+70,+51){$L$} \put(+90,+51){$M$}
\put(+70,+2){$P$} \put(+90,+2){$S$}

\put(+50,+0){\vector(0,+1){60}}  \put(+50,+60){$\phi $}

\put(+0,+50){\line(+1,0){100}}   \put(+95,+40){$A$}
\put(+0,+10){\line(+1,0){100}}   \put(+95,+20){$B$}
\put(+50,+50){$+2\pi $}           \put(+40,+5){$0 $}

\put(+10,+10){\line(0,+1){40}}   \put(+30,+10){\line(0,+1){40}}
\put(+70,+10){\line(0,+1){40}}   \put(+90,+10){\line(0,+1){40}}

\end{picture}

\begin{center}
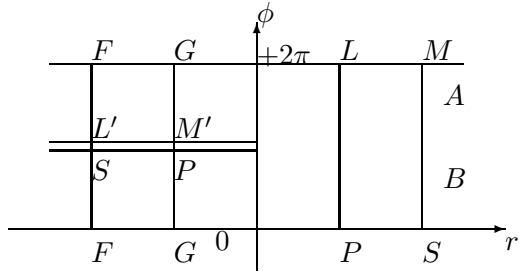
FIG. 31. Yet else one $(\phi, r)$ region
\end{center}

\vspace{5mm}

It is self-evident that everything said about
$\xi$-model is  suitable for another spinor model $\eta$
as well.

\section{Conclusion}

The results obtained for  3-space with $(x,y,z)$ coordinates
 should  be extended to  Minkowski 4-space with coordinates $(t,x.y,z)$.
Mathematically it  means the  use of relativistic $SL(2.C)$ spinors instead of
non-relativistic  $SU(2)$ spinors.

Domains  of curvilinear coordinates associated with spinor space
 can be used to examine  possible  quantum mechanical
manifestation of the  spinor structure
both in non-relativistic and relativistic theories.
 To this end, one should  specially look at  analytical
properties of the  known solutions
of the Schr\"{o}dinger and Dirac equations in various
coordinates.

\label{last}

\end{document}